\documentclass[conference]{IEEEtran}

\IEEEoverridecommandlockouts
\usepackage{amsmath,amsfonts}
\usepackage{algorithmic}
\usepackage{array}
\usepackage[caption=false,font=normalsize,labelfont=sf,textfont=sf]{subfig}
\usepackage{textcomp}
\usepackage{stfloats}
\usepackage{url}
\usepackage{verbatim}
\usepackage{graphicx}

\def\BibTeX{{\rm B\kern-.05em{\sc i\kern-.025em b}\kern-.08em
    T\kern-.1667em\lower.7ex\hbox{E}\kern-.125emX}}

\usepackage{balance}

\usepackage[linesnumbered,ruled,vlined]{algorithm2e}
\SetKwInput{KwInput}{Input}                % Set the Input
\SetKwInput{KwOutput}{Output}              % set the Output
\SetCommentSty{mycommfont}

\usepackage{multirow} %% for multi-rows
\usepackage{booktabs}
\usepackage{color,soul} % for highlight \hl
\usepackage{bm} % for math symbold to be bold

\usepackage{dutchcal}

\newcommand\norm[1]{\lVert#1\rVert}

\usepackage{outlines}

\usepackage[table]{xcolor}
\definecolor{Gray}{gray}{0.9}
\definecolor{Yellow}{rgb}{1.0, 0.97, 0.7}

\usepackage{tablefootnote}

\usepackage{comment}

\usepackage{enumitem}

\usepackage{scalerel}
\usepackage{tikz}
\usetikzlibrary{svg.path}

\definecolor{orcidlogocol}{HTML}{A6CE39}
\tikzset{
    orcidlogo/.pic={
        \fill[orcidlogocol] svg{M256,128c0,70.7-57.3,128-128,128C57.3,256,0,198.7,0,128C0,57.3,57.3,0,128,0C198.7,0,256,57.3,256,128z};
        \fill[white] svg{M86.3,186.2H70.9V79.1h15.4v48.4V186.2z}
        svg{M108.9,79.1h41.6c39.6,0,57,28.3,57,53.6c0,27.5-21.5,53.6-56.8,53.6h-41.8V79.1z M124.3,172.4h24.5c34.9,0,42.9-26.5,42.9-39.7c0-21.5-13.7-39.7-43.7-39.7h-23.7V172.4z}
        svg{M88.7,56.8c0,5.5-4.5,10.1-10.1,10.1c-5.6,0-10.1-4.6-10.1-10.1c0-5.6,4.5-10.1,10.1-10.1C84.2,46.7,88.7,51.3,88.7,56.8z};
    }
}

\newcommand\orcidicon[1]{\href{https://orcid.org/#1}{\mbox{\scalerel*{
                \begin{tikzpicture}[yscale=-1,transform shape]
                \pic{orcidlogo};
                \end{tikzpicture}
            }{|}}}}

\usepackage{hyperref} %<--- Load after everything else

\begin{document}

\title{Speed-up of Data Analysis with Kernel Trick \\in Encrypted Domain}

% \author{
%     \IEEEauthorblockN{Anonymous}
%     % \IEEEauthorblockA{\textit{Anonymous Affiliation} \\
%     % Anonymous City, Country \\
%     % email@domain.com}
% }

\author{
Joon Soo Yoo, Baek Kyung Song, Tae Min Ahn, Ji Won Heo, Ji Won Yoon \\
\textit{Korea University} \\
Seoul, South Korea \\
\{sandiegojs, baekkyung777, xoals3563, hjw4, jiwon\_yoon\}@korea.ac.kr
}

% \and
% \and
% \and
% \and
% \and

\maketitle
% remove later
\thispagestyle{plain}
\pagestyle{plain}

\begin{abstract}

% Homomorphic encryption (HE) is pivotal for secure computation on encrypted data, crucial in privacy-preserving data analysis. However, efficiently processing high-dimensional data in HE, especially for machine learning and statistical (ML/STAT) algorithms, poses a challenge. In this paper, we present $\phi$-ZER, an effective acceleration method using the kernel trick for HE schemes, enhancing time performance in ML/STAT algorithms within encrypted domains. This technique, independent of underlying HE mechanisms and complementing existing optimizations, notably reduces costly HE multiplications, offering near constant time complexity relative to data dimension. Aimed at accessibility, $\phi$-ZER is tailored for data scientists and developers with limited cryptography background, facilitating advanced data analysis in secure environments.

Homomorphic encryption (HE) is pivotal for secure computation on encrypted data, crucial in privacy-preserving data analysis. However, efficiently processing high-dimensional data in HE, especially for machine learning and statistical (ML/STAT) algorithms, poses a challenge. In this paper, we present an effective acceleration method using the kernel method for HE schemes, enhancing time performance in ML/STAT algorithms within encrypted domains. This technique, independent of underlying HE mechanisms and complementing existing optimizations, notably reduces costly HE multiplications, offering near constant time complexity relative to data dimension. Aimed at accessibility, this method is tailored for data scientists and developers with limited cryptography background, facilitating advanced data analysis in secure environments.

\end{abstract}

\begin{IEEEkeywords}
Homomorphic Encryption, Kernel Method, Privacy-Preserving Machine Learning, Privacy-Preserving Statistical Analysis, High-dimensional Data Analysis
\end{IEEEkeywords}

\section{Introduction}
Homomorphic encryption (HE) enables computations on encrypted data, providing quantum-resistant security in client-server models. Since its introduction by Gentry in 2009~\cite{gentry2009fully}, HE has rapidly evolved towards practical applications, offering privacy-preserving solutions in various data-intensive fields such as healthcare and finance.

Despite these advancements, applying HE to complex nonlinear data analysis poses significant challenges due to the computational demands of internal nonlinear functions. Although basic models such as linear regression are well-suited for HE, extending this to more complex algorithms, like those seen in logistic regression models~\cite{kim2018logistic, chen2018logistic, aono2016scalable, crockett2020low}, often results in oversimplified linear or quasi-linear approaches. This limitation narrows the scope of HE's applicability to more intricate models.

Furthermore, advanced data analysis algorithms resist simple linearization, shifting the focus of research to secure inference rather than encrypted training. Prominent models like nGraph-HE2~\cite{boemer2019ngraph}, LoLa~\cite{brutzkus2019low}, and CryptoNets~\cite{gilad2016cryptonets} exemplify this trend, emphasizing inference over training, with reported latencies on benchmarks like the MNIST dataset of 2.05 seconds for nGraph-HE2, 2.2 seconds for LoLa, and 250 seconds for CryptoNets. It should be noted that these models are only used for testing and do not encrypt trained parameters, relying instead on relatively inexpensive plaintext-ciphertext multiplications, which contributes to their reported time performance.

Since all of these problems arise from the time-consuming nature of operations performed in HE schemes,  much literature focuses on optimizing the performance level of FHE by hardware acceleration or adding new functionality in the internal primitive. Jung et al.~\cite{jung2021over} propose a memory-centric optimization technique that reorders primary functions in bootstrapping to apply massive parallelism; they use GPUs for parallel computation and achieve 100 times faster bootstrapping in one of the state-of-the-art FHE schemes, CKKS~\cite{cheon2017homomorphic}. Likewise, \cite{jung2021accelerating} illustrates the extraction of a similar structure of crucial functions in HE multiplication; they use GPUs to improve time performance by 4.05 for CKKS multiplication. Chillotti et al.~\cite{chillotti2021programmable} introduces the new concept of programmable bootstrapping (PBS) in TFHE library~\cite{chillotti2016faster, chillotti2020tfhe} which can accelerate neural networks by performing PBS in non-linear activation function.

Our approach to optimization is \emph{fundamentally different} from existing techniques, which primarily focus on hardware acceleration or improving cryptographic algorithms within specific schemes or libraries. Unlike these approaches, our optimizer operates \emph{independently} of underlying cryptographic schemes or libraries, and works at a higher software (SW) level. As a result, it does not compete with other optimization techniques and can synergistically amplify their effects to improve time performance.

\begin{comment}
We take a completely \emph{different} approach than the current optimization technique. Our novelty is that the number of data $n$ and dimension $d$ should always be \emph{known} to the server regardless of the evaluation being performed over the encrypted domain. That is, the cloud has to know the information about the client's data size. Otherwise, the cloud has to run over all possible choices of the parameters $n$ and $d$, which is practically \emph{impossible}. 

Our approach, the kernel optimizer, requires only the known or kernel parameters. The cloud uses the kernel parameters to calculate the time complexities of the two evaluation circuits, namely, general and kernel circuits. Then, the cloud can select the circuit based on the time complexities for better performance.

Moreover, our optimizer works in the upper or SW level as opposed to the current optimization technique targeted at the internal or HW level. Therefore, using our optimizer at the \emph{upper} level with other internal HE accelerators can amplify the boosting effect improving the time performance significantly. 
\end{comment}

Moreover, our approach does not necessitate extensive knowledge of cryptographic design. By utilizing our technique, significant speed-ups in homomorphic circuits, particularly for machine learning and statistical algorithms, can be achieved with relative ease by data scientists and developers. Our approach requires only a \emph{beginner-level understanding of cryptography}, yet yields remarkable improvements in performance that are difficult to match, especially for high-dimensional ML/STAT analysis.

Furthermore, our approach holds great promise for enabling \emph{training} in the encrypted domain, where traditional most advanced machine learning methods struggle to perform. We demonstrate the effectiveness of our technique by applying it to various machine learning algorithms, including support vector machine (SVM), $k$-means clustering, and $k$-nearest neighbor algorithms. In classical implementations of these algorithms in the encrypted domain, the practical execution time is often prohibitive due to the high-dimensional data requiring significant computation. However, our approach provides a nearly \emph{dimensionless} property, where the required computation for high-dimensional data remains nearly the same as that for low-dimensional data, allowing for efficient training even in the encrypted domain.

Our optimizer is based on the kernel method~\cite{zaki2014data}, a widely-used technique in machine learning for identifying nonlinear relationships between attributes. The kernel method involves applying a function $\phi$ that maps the original input space to a higher-dimensional inner product space. This enables the use of modified versions of machine learning algorithms expressed solely in terms of kernel elements or inner products of data points. By leveraging this approach, our optimizer enables the efficient training of machine learning algorithms even in the presence of high-dimensional data, providing a powerful tool for encrypted machine learning.

Our findings suggest that the kernel trick can have a significant impact on HE regarding execution time. This is due to the substantial time performance gap between addition and multiplication operations in HE. In HE schemes, the design of multiplication is much more complex. For example, in TFHE, the multiplication to addition is about 21, given the 16-bit input size for Boolean evaluation. In addition, BGV-like schemes (BGV, BFV, CKKS) require additional steps followed by multiplication of ciphertexts such as relinearization and modulus-switching. On contrary, the latency of plain multiplication is approximately the same as that of plain addition ~\cite{guide2011intel}. 

The kernel method can effectively reduce the number of heavy multiplication operations required, as it transforms the original input space to an inner product space where kernel elements are used for algorithm evaluation. Since the inner product between data points is pre-evaluated in the preprocessing stage, and since the inner product involves the calculation of $d$ (dimension size) multiplications, much of the multiplication in the algorithm is reduced in this process. This structure can lead to significant performance improvements for some ML/STAT algorithms, such as $k$-means and total variance.

To showcase the applicability of the kernel method to encrypted data, we performed SVM classification as a preliminary example on a subset of the MNIST dataset~\cite{deng2012mnist} using the CKKS encryption scheme, which is implemented in the open-source OpenFHE library~\cite{al2022openfhe}. Specifically, we obtained an estimation\footnote{$\lambda = 128$ represents the security parameter, $N=2^{16}$ signifies the dimension of the polynomial ring, $\Delta = 2^{50}$ is the scaling factor, and $L=110$ stands for the circuit depth (Check~\cite{cheon2017homomorphic}.)} of 38.18 hours and 509.91 seconds for SVM's general and kernel methods, respectively. This demonstrates a performance increase of approximately 269 times for the kernel method compared to the classical approach.

We summarize our contributions as follows:

\begin{itemize}[leftmargin=0.35cm]
    \item \textbf{Universal Applicability.} Introduced the linear kernel trick to the HE domain, where our proposed method works \emph{independently}, regardless of underlying HE schemes or libraries. The kernel method can \emph{synergistically} combine with any of the underlying optimization techniques to boost  performance. 
    \item \textbf{Dimensionless Efficiency.} Demonstrated near-constant time complexity across ML/STAT algorithms (classification, clustering, dimension reduction), leading to significant execution time reductions, especially for high-dimensional data.
    \item \textbf{Enhanced Training Potential.} Shown potential for significantly improved ML training in the HE domain where current HE training models struggle.
    % \item \textbf{Systematic Optimization.} Developed $\phi$-ZER, a systematic framework for choosing optimal circuit evaluation (general or kernel) based on time complexity.
    \item \textbf{User-Friendly Approach.} Easily accessible for data scientists and developers with limited knowledge of cryptography. 

\end{itemize}
\section{Preliminaries}

\subsection{Homomorphic Encryption (HE)} 

Homomorphic encryption (HE) is a quantum-resistant encryption scheme that includes a unique evaluation step, allowing computations on encrypted data. The main steps for a general two-party (\textsf{client-server} model) symmetric HE scheme are as follows, with the \textsf{client} performing:

\begin{itemize}[leftmargin=0.35cm]

\item \textsf{KeyGen($\lambda$)}: Given a security parameter $\lambda$, the algorithm outputs a secret key $\mathsf{sk}$ and an evaluation key $\mathsf{evk}$.

\item \textsf{Enc($m$, $\mathsf{sk}$)}: Given a message $m \in M$ from the message space $M$, the encryption algorithm uses the secret key $\mathsf{sk}$ to generate a ciphertext $\mathsf{ct}$.

\item \textsf{Dec($\mathsf{ct}$, $\mathsf{sk}$)}: Given a ciphertext $\mathsf{ct}$ encrypted under the secret key $\mathsf{sk}$, the decryption algorithm outputs the original message $m \in M$.

\end{itemize}

A distinguishing feature of HE, compared to other encryption schemes, is the evaluation step, \textsf{Eval}, which computes on encrypted messages and is performed by the \textsf{server}. 

\begin{itemize}[leftmargin=0.35cm]

\item \textsf{Eval}($\mathsf{ct}_1, \cdots, \mathsf{ct}_k$, $\mathsf{evk}$; $\psi$): Suppose a function $\psi : M^k \rightarrow M$ is to be performed over messages $m_1, \dots, m_k$. The evaluation algorithm takes as input the ciphertexts $\mathsf{ct}_1, \cdots, \mathsf{ct}_k$, each corresponding to $m_1, \cdots, m_k$ encrypted under the same $\mathsf{sk}$, and uses $\mathsf{evk}$ to generate a new ciphertext $\mathsf{ct}'$ such that $\mathsf{Dec}(\mathsf{ct}', \mathsf{sk}) = \psi(m_1, \cdots, m_k)$.

\end{itemize}

% Homomorphic encryption consists of 4 steps: key generation, encryption, decryption, and evaluation. 
% %
% \begin{itemize}[leftmargin=0.35cm]
%     \item \textbf{Key generation}: given a security parameter $\lambda$, the key generation algorithm outputs a pair of keys: a secret key $sk$ for the private key variant, and both a secret key $sk$ and a public key $pk$ for the public key variant. Additionally, an evaluation key $evk$ is produced.
%     \item \textbf{Encryption}: encryption algorithm is a probabilistic algorithm that takes in a message $m \in \mathsf{M}$ and a secret (or public) key $sk$ as input and outputs a random ciphertext $c \in \mathcal{C}$. The correctness property ensures that the ciphertext $c$ is decrypted to the message $m$ deterministically.
%     \item \textbf{Decryption}: given a valid ciphertext $c$, the decryption algorithm takes in the ciphertext $c$ and secret key $sk$ to output a message $m$.
%     \item \textbf{Evaluation}: Suppose a function $\psi : \mathsf{M}^k \rightarrow \mathsf{M}$ is performed over the messages $m_1, \dots, m_k$. Then, the evaluation algorithm takes in $c_1, \dots, c_k$ corresponding to $m_1, \dots, m_k$ and $evk$ to output $c^*$ such that $\textsf{Dec}(c^*) = \psi(m_1, \dots, m_k)$.  
% \end{itemize}
% %

\noindent \textbf{Fully or Leveled HE.} The most promising HE schemes rely on lattice-based problem---learning with errors (LWE) which Regev~\cite{regev2009lattices} proposed in 2005, followed by its ring variants by Stehlé~\cite{stehle2009efficient} in 2009. LWE-based HE uses noise for security; however, the noise is accumulated for each evaluation of ciphertext. When the noise of ciphertext exceeds a certain limit, the decryption of ciphertext does not guarantee the correct result. Gentry~\cite{gentry2009fully} introduces a bootstrapping technique that periodically reduces the noise of the ciphertext for an unlimited number of evaluations, i.e., \emph{fully homomorphic}. However, such bootstrapping is a costly operation. Thus many practical algorithms bypass bootstrapping and instead use \emph{leveled homomorphic encryption}, or LHE---the depth of the circuit is pre-determined; it uses just enough parameter size for the circuit to evaluate without bootstrapping. 

\noindent \textbf{Arithmetic or Boolean.} There are two primary branches of FHE: arithmetic and Boolean-based. (1) The \emph{arithmetic HE}---BGV~\cite{brakerski2014leveled} family---uses only addition and multiplication. Thus, it has to approximate nonlinear operations with addition and multiplication. Also, it generally uses the LHE scheme and has a faster evaluation. B/FV~\cite{fan2012somewhat} and CKKS~\cite{cheon2017homomorphic} are representative examples. 
(2) \emph{Boolean-based HE} is the GSW~\cite{gentry2013homomorphic} family---FHEW~\cite{ducas2015fhew} and TFHE~\cite{chillotti2020tfhe}, which provides fast-bootstrapping gates such as XOR and AND gates. Typically, it is slower than arithmetic HE; however, it has more functionalities other than the usual arithmetics.

% \noindent \textbf{Worst-case Design Principle.} It is important to note that every circuit---either arithmetic or Boolean---should be designed in the worst-case scenario. Matching the convergence condition or selecting the possible shortcut for the circuit evaluation is impossible since we compute functions with the encrypted values. For example, the if/else condition requires both circumstances to be executed due to the encrypted input.

% \noindent \textbf{Costly HE Multiplication.} All FHE schemes have expensive multiplication operations. The addition operation involves a simple vector addition; however, the multiplication operation entails a series of supplementary measures, thus, is time-consuming. For instance, the CKKS scheme requires relinearization and modulus switching procedures after two ciphertext multiplications. Moreover, the circuit depth is pre-determined by the multiplication number due to their significant noise increase and structural features.

\subsection{Kernel Method}

In data analysis, the kernel method is a technique that facilitates the transformation $\phi: X \rightarrow F$ from the input space $X$ to a high-dimensional feature space $F$, enabling the separation of data in $F$ that is not linearly separable in the original space $X$. The kernel function $K(\mathbf{x_i}, \mathbf{x_j}) = \phi(\mathbf{x_i}) \cdot \phi(\mathbf{x_j})$ computes the inner product of $\mathbf{x_i}$ and $\mathbf{x_j}$ in $F$ without explicitly mapping the data to the high-dimensional space $F$. The kernel matrix $\mathbf{K}$ is a square symmetric matrix of size $n \times n$ that contains all pair-wise inner products of $\phi(\mathbf{x_i})$ and $\phi(\mathbf{x_j})$, i.e., $K(\mathbf{x_i}, \mathbf{x_j})$, where the data matrix $\mathbf{D}$ contains $n$ data points $\mathbf{x_i}$ each with dimension $d$. For more details, refer to \cite{scholkopf2002learning}.

\section{Kernel Effect in Homomorphic Encryption}
\label{sec:ker_effect}

% This section illustrates the reasoning behind the effectiveness of the kernel in the homomorphic encryption domain. In particular, we provide two kernel properties that can improve the time performance of the homomorphic circuit:

% \begin{itemize}[leftmargin=0.35cm]
%     \item \textbf{(P1)} Parallel computation of kernel function $K(\cdot, \cdot)$
%     \item \textbf{(P2)} Dimensionless: constant time complexity w.r.t. $d$
%     % \item \textbf{(P3)} Different input sizes for inverse matrix operation
% \end{itemize}

% Moreover, we provide a concrete example---$k-$means---to show that the kernel properties can effectively improve homomorphic circuit evaluation. 

\subsection{Why is Kernel More Effective in HE than Plain Domain?}
The kernel method can greatly benefit from the structural difference in HE between addition and multiplication operations. Specifically, the main reason is that \emph{the kernel method avoids complex-designed HE multiplication and uses almost-zero cost additions}. 

\noindent \textbf{Time-consuming HE Multiplication.} Homomorphic multiplication is significantly more complex than addition, unlike plain multiplication, which has a similar time performance to addition. For example, in the BGV family of the homomorphic encryption scheme, the multiplication of two ciphertexts $\mathbf{ct} = (\mathsf{ct_0}, \mathsf{ct_1})$ and $\mathbf{ct'} = (\mathsf{ct_0'}, \mathsf{ct_1'}) \in \mathcal{R}_q^2$ is defined as:
\begin{equation*}
    \mathbf{ct}_{\text{mult}} = (\mathsf{ct_0} \cdot \mathsf{ct_0'}, \mathsf{ct_0} \cdot \mathsf{ct_1'} + \mathsf{ct_0'} \cdot \mathsf{ct_1}, \mathsf{ct_1} \cdot \mathsf{ct_1'})   
\end{equation*}
where $\mathcal{R}_q = \mathbb{Z}_q[X]/(X^N + 1)$ is the polynomial ring. The size of the ciphertext $\mathbf{ct}_{\text{mult}}$ grows after multiplication, requiring an additional process called \emph{relinearization} to reduce the ciphertext to its normal form. In the CKKS scheme, a \emph{rescaling} procedure is used to maintain a constant scale. On the other hand, HE addition is merely vector addition of polynomial ring elements $\mathcal{R}_q$. Thus, the performance disparity between addition and multiplication in HE is substantial compared to that in plain operations (see Table~\ref{tab:add_mult_ratio}).

% Table~\ref{tab:add_mult_ratio} shows the ratio of addition and multiplication execution times for plain and various ciphertext domains.

%

% We use parameters $N=2^{16} $ for CKKS scheme. Also, $N=2^{16}$ are used for measuring B/FV operations. Lastly, we use $16-$bit input for TFHE evaluation

% ctx vs ptx: add and mult with ratios
\begin{table}[htb!]
\caption{Comparison of Execution Times and Ratios for Addition and Multiplication in Plain and HE Domains.}
\begin{minipage}{\linewidth}
\begin{center}
\resizebox{\linewidth}{!} { 
\begin{tabular}{c  cccc}
\toprule
 & Plain\tablefootnote{We obtained the average execution time of plaintext additions and multiplications by measuring their runtime 1000 times each.} & TFHE\tablefootnote{$\lambda = 128, l=16$} & CKKS\tablefootnote{$\lambda = 128, N=2^{16}, \Delta = 2^{50}, L = 50$} & B/FV\tablefootnote{$\lambda = 128, L=20, n=2^{15}, \log_2(q)=780$} \\
 
\midrule

Addition & $3.39 (ns)$ & $ 1.06 (s)$ & $24.85 (ms)$ & $ 1.81 (ms) $ \\

Multiplication & $3.56 (ns)$ & $ 22.95 (s) $ & $920.75 (ms)$ & $ 284.62 (ms) $ \\

\midrule

\rowcolor{Gray}
Ratio & $1.05$ & $21.65$ & $37.04$ & $156.73$ \\

\bottomrule

\end{tabular}
}
\end{center}
\end{minipage}
\label{tab:add_mult_ratio}
\end{table}

% \subsubsection{Limited Multiplicative Depth} Furthermore, HE multiplication is limited to a certain degree within leveled homomorphic encryption scheme. In leveled BGV family, the depth of the circuit is pre-determined due to the noise accumulation and modulus reduction. Specifically, multiplication increases noise significantly compared to addition. Also, in the CKKS scheme, multiplication induces modulus reduction of the ciphertext space; thus, the limited ciphertext space prevents further operation. Conversely, the addition operation is approximately zero-cost regarding circuit depth. In principle, it is recommended to use less multiplicative depths for homomorphic circuit efficiency; typically, more multiplicative depths require larger parameters, resulting in performance decline. 

% In short, the kernel method can dramatically improve performance using additions to replace expensive multiplications. We examine the kernel properties that can bypass heavy multiplications for homomorphic circuit acceleration. 

\subsection{Two Kernel Properties for Acceleration}
% \noindent\textbf{P1: Parallel Computation of Kernel Function.} The kernel trick uses the Hilbert space, where the elements of the space are inner products of feature vectors: $K(\mathbf{x}_i, \mathbf{x}_j) = \phi(\mathbf{x}_i)^T \phi(\mathbf{x}_j)$. In particular, we use the linear kernel trick that uses identity mapping $\phi: \mathbf{x} \mapsto \mathbf{x}$. Therefore, the linear kernel elements are simply the dot products of the vectors: $K(\mathbf{x}_i, \mathbf{x}_j) = \mathbf{x}_i^T \mathbf{x}_j$. This identical structure of computing kernel elements induces \emph{parallel computation}. That is, we can assign processors to evaluate each kernel element for concurrency. Fig.~\ref{fig:ker_prop_1} illustrates the concept of parallel computation of kernel elements. 

\begin{figure}[!htb]
    \centering
    \includegraphics[width=0.48\textwidth]{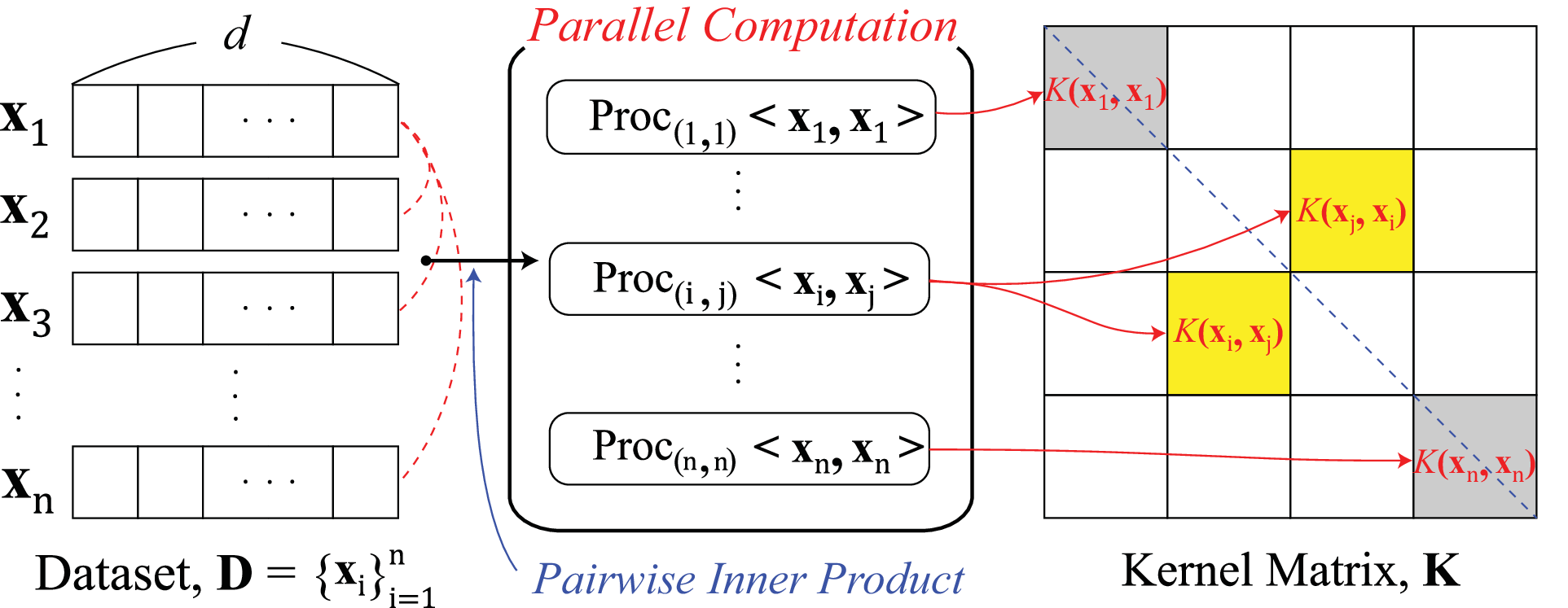}
    % \caption{(\textbf{P1}) \emph{Parallel computation structure for evaluating linear kernel elements}. We can \emph{pre-compute} the expensive homomorphic multiplication \emph{parallel} in the opening stage for high performance. For each evaluation of kernel element $K(\mathbf{x}_i, \mathbf{x}_j) = \langle \mathbf{x}_i, \mathbf{x}_j \rangle$, we can assign a processor for concurrent computation over the kernel matrix \textbf{K}. By symmetry of \textbf{K}, it requires $\frac{n(n+1)}{2} \leq n^2$ processors for parallel computation.}
    \caption{(\textbf{P1}) \emph{Parallel computation structure for evaluating kernel elements}. Expensive HE multiplications can be pre-computed in parallel during the initial stage for enhanced performance. Each kernel element evaluation can be assigned to a separate processor, enabling concurrent computation over the kernel matrix $\mathbf{K}$.}
    \label{fig:ker_prop_1}
\end{figure}

\noindent \textbf{P1: Parallel Computation of Kernel Function.} One property of kernel evaluation is the ability to induce parallel structures in the algorithm, such as computing kernel elements $K(\mathbf{x_i}, \mathbf{x_j}) = \phi(\mathbf{x_i}) \cdot \phi(\mathbf{x_j})$. Since evaluating kernel elements involves performing the \emph{same dot product structure} over input vectors of the same size, HE can benefit from parallel computation of these kernel elements.
For instance, in the basic HE model, (1) the \textsf{server} can perform concurrent computation of kernel elements during the \emph{pre-processing stage}. (2) Alternatively, the \textsf{client} can compute the kernel elements and send the encrypted kernels to the \textsf{server} as alternative inputs, replacing the original encrypted data for further computation. Either way, the \textsf{server} can \emph{bypass the heavy multiplication in the pre-processing stage}, thereby accelerating the overall kernel evaluation process (see Fig.~\ref{fig:ker_prop_1}).

\begin{figure}[!htb]
    \centering
    \includegraphics[width=0.48\textwidth]{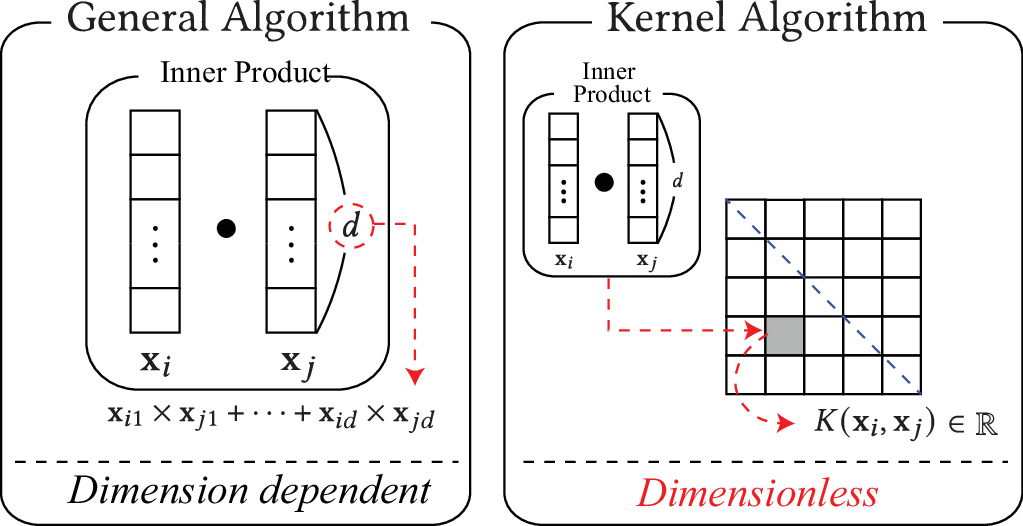}
    \caption{(\textbf{P2}) \emph{Dimensionless---constant time complexity w.r.t. dimension}. While general ML/STAT algorithms require the computation of inner products involving costly multiplications, the kernel method bypasses these dot products by utilizing kernel elements or scalars.}
    \label{fig:ker_prop_2}
\end{figure}

\noindent\textbf{P2: Dimensionless---Constant Time Complexity with respect to Dimension.} A key feature of the kernel method in HE schemes is its \emph{dimensionless} property. Generally, evaluating certain ML/STAT algorithms requires numerous inner products of vectors, each involving $d$ multiplications. Given the high computational cost of HE multiplications, this results in significant performance degradation. However, by employing the kernel method, we can circumvent the need for these dot products. Consequently, dimension-dependent computations are avoided during kernel evaluation, leading to a total time complexity that is \emph{constant} with respect to the data dimension $d$, unlike general algorithms, which have a linear dependency on $d$ (see Fig.~\ref{fig:ker_prop_2}). 

\subsection{\texorpdfstring{Example: $k$-means Algorithm}{Example: k-means Algorithm}}
% \subsection{Example: $k-$means Algorithm}

\begin{figure}[!htb]
    \centering
    \includegraphics[width=0.48\textwidth]{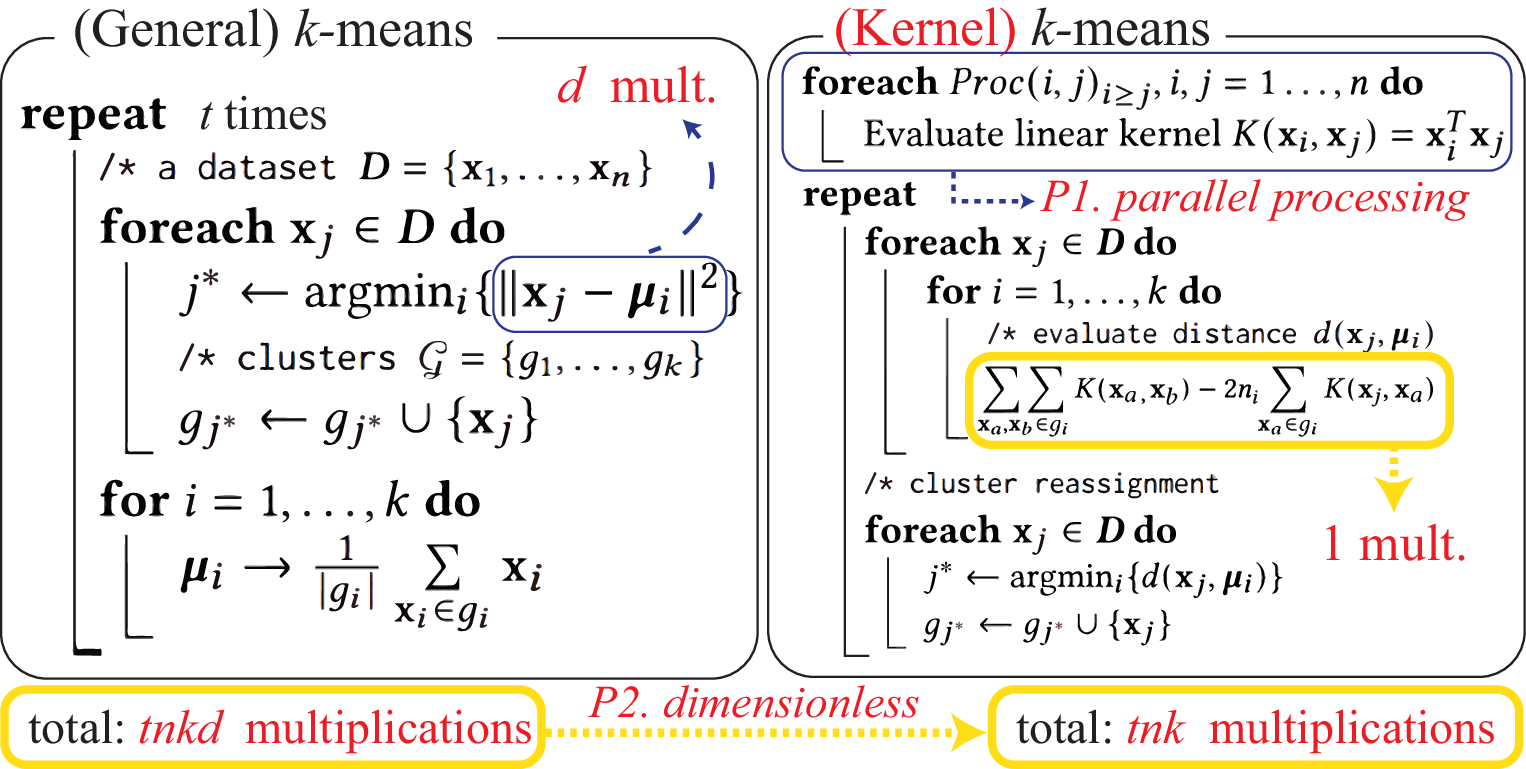}
    \caption{Example of the $k-$means algorithm. The kernel method lessens the multiplication number by approximately a factor of $d$ (P1). The inner products or the kernel element is computed parallel from the initial stage (P2). The total time complexity is nearly constant with respect to dimension $d$.}
    \label{fig:kmeans_gen_ker}
\end{figure}

The $k-$means algorithm is an iterative process for finding optimal clusters. The algorithm deals with (1) cluster reassignment and (2) centroid update (see Fig.~\ref{fig:kmeans_gen_ker}). The bottleneck is the cluster reassignment process that computes the distance $d(\mathbf{x}_j, \bm{\mu}_i)$ from each centroid $\bm{\mu}_i$. Since evaluating the distance requires a dot product of the deviation from the centroid, $d$ multiplications are required for calculating each distance. Assuming that the algorithm converges in $t$ iteration, the total number of multiplications is $tnkd$. 
%--- kmeans general vs kernel figure
% \begin{figure*}[!htb]
%     \centering
%     \includegraphics[width=0.9\textwidth]{figures/kmeans2.eps}
%     \caption{k-means}
%     \label{fig:kmeans_gen_ker}
% \end{figure*}

Using the kernel method and its properties, we can significantly improve the time performance of the circuit. First, we express the distance in terms of the kernel elements:
\begin{equation}
     d(\mathbf{x}_j, \bm{\mu}_i) = \mathop{\sum\sum}_{\mathbf{x}_a, \mathbf{x}_b \in g_i} K(\mathbf{x}_a, \mathbf{x}_b) - 2n_i \sum\limits_{\mathbf{x}_a \in g_i} K(\mathbf{x}_j, \mathbf{x}_a) 
\label{eq:kmeans_eq_kernel}
\end{equation}
where $n_i$ is the number of elements in the cluster $g_i \in \{g_1, \dots, g_k\}$. We verify that only one multiplication is performed in Eq.~(\ref{eq:kmeans_eq_kernel}), which reduces the multiplications by a factor of $d$ from the general distance calculation. Therefore, the total number of multiplications in the kernel $k-$means is $tnk$ (P2, \emph{dimensionless} property), whereas $tnkd$ in the general method. Note that we can evaluate the kernel elements in parallel at the algorithm's beginning, compensating for the $d$ multiplication time (P1, \emph{parallel} structure). 

% Moreover, the kernel $k-$means method likely increases the additions from the general method. We can check that the additions in $d(\mathbf{x}_j, \bm{\mu}_i)$ has changed to $O(|g_i|^2)$ from $O(d)$ in the general distance. The total time complexity of addition is $O(tn^3)$ for the kernel method, whereas $O(tnkd)$ for the general method. Hence, the kernel method evaluates more additions depending on the parameter set. 

\begin{figure}[htb]
    \centering
    \subfloat[$(n=10, d=784)$]{\includegraphics[width=0.235\textwidth]{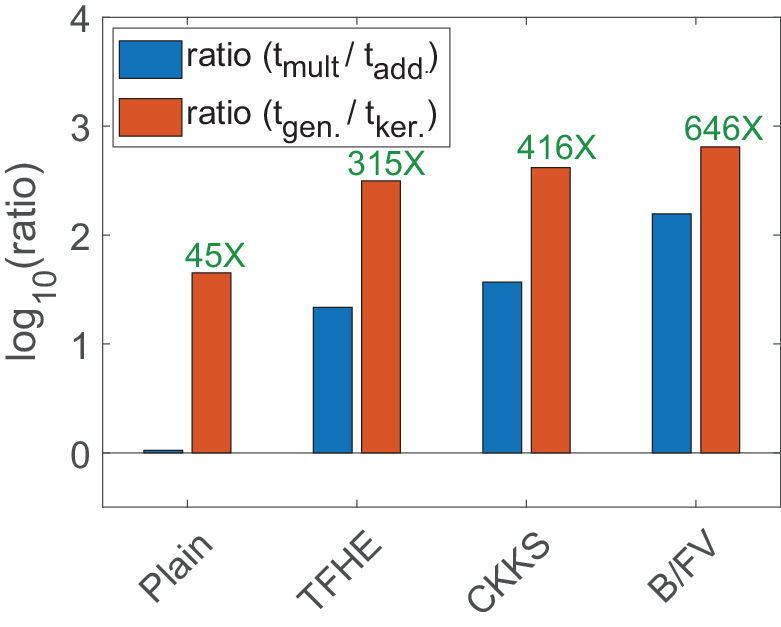}
    \label{fig:ker_eff_param_1}}
    \hfill
    \subfloat[$(n=100, d=784)$]{\includegraphics[width=0.235\textwidth]{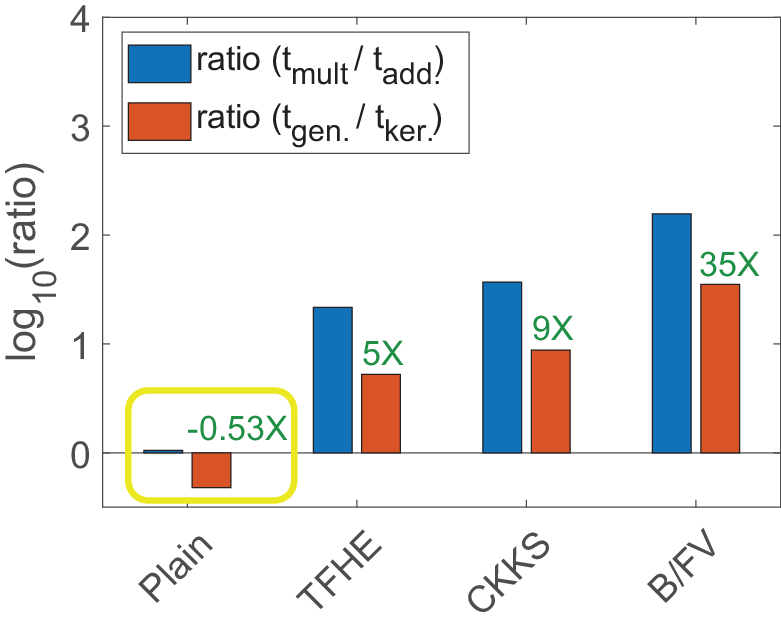}
    \label{fig:ker_eff_param_2}}
    \caption{$k$-means simulation: This figure demonstrates the effectiveness of the kernel method in $k$-means across different domains—plain, TFHE, CKKS, and B/FV. The parameters were fixed at $k=3$ and $t=10$, and the log ratio of general time ($t_{gen}$) to kernel time ($t_{ker}$) was compared. 
    % The result (a) shows that the more significant the time difference gap, the more efficient the kernel method applies. The result (b) demonstrates that the impact of the kernel method on the performance varies with changes in the parameters of $n$ and $d$. We note that the plain kernel demonstrates the least benefit from the kernelization.
    }
    \label{fig:k_means_ratio}
\end{figure}

\noindent \textbf{Kernel Method---More Effective in the Encrypted Domain: $k-$means Simulation.} Based on algorithms in Fig.~\ref{fig:kmeans_gen_ker}, we simulate the $k$-means algorithm within different types of domains---plain, TFHE, CKKS, and B/FV---to demonstrate the effectiveness of the kernel method in the encrypted domain (see Fig.~\ref{fig:k_means_ratio}). We count the total number of additions and multiplications in $k-$means; based on the HE parameter set and execution time in Table~\ref{tab:add_mult_ratio}, we compute the total simulation time with fixed parameters $t=10, k=3$.
%

% \begin{figure}[htb]
%     \centering
%     \begin{subfigure}[t]{0.235\textwidth}
%         % \centering
%         \includegraphics[width=\textwidth]{figures/ker_eff_param_1.eps}
%         \caption{$(n=10, d=784)$}
%         \label{fig:ker_eff_param_1}
%     \end{subfigure}
%     \hfill
%     \begin{subfigure}[t]{0.235\textwidth}
%         % \centering
%         \includegraphics[width=\textwidth]{figures/ker_eff_param_2.eps}
%         \caption{$(n=100, d=784)$}
%         \label{fig:ker_eff_param_2}
%     \end{subfigure} 
%     \caption{$k-$means simulation: the figure shows the effectiveness of the kernel method in $k-$means concerning different domains---plain, TFHE, CKKS, and B/FV. We fixed the parameters $k=3, t=10$ and compared their log ratio (general time $t_{gen}$ /  kernel time $t_{ker}$). 
%     % The result (a) shows that the more significant the time difference gap, the more efficient the kernel method applies. The result (b) demonstrates that the impact of the kernel method on the performance varies with changes in the parameters of $n$ and $d$. We note that the plain kernel demonstrates the least benefit from the kernelization.
%     }
%     \label{fig:k_means_ratio}
% \end{figure}

The result indicates that the kernel method is more effective in the encrypted domain. Fig.~\ref{fig:ker_eff_param_1} shows that the kernel method has a significant speed-up when the time ratio of addition and multiplication is large. For instance, B/FV has a maximum speed-up of $645$ times, whereas plain kernel has $45$ times (CKKS: $416$ times, TFHE: $315$ times). Moreover, Fig.~\ref{fig:ker_eff_param_2} demonstrates that depending on the parameter set, the kernel effect on the time performance can differ. For instance, B/FV has a speed-up of $35$ times in Fig.~\ref{fig:ker_eff_param_2}.

Furthermore, the kernel can even have a negative impact on the execution time. For example, the plain kernel in Fig.~\ref{fig:ker_eff_param_2} conversely decreases the total execution time by $53$ percent. This highlights the sensitivity of the plain kernel's performance to variations. We provide actual experimental results regarding the kernel effect in Section~\ref{section:result}.

\subsection{Benefits of the Kernel Method}

\noindent\textbf{Software-Level Optimization---Synergistic with Any HE Schemes or Libraries.} The kernel method is \emph{synergistic} with any HE scheme, amplifying the speed-up achieved by HE hardware accelerators since it operates at the software level. For instance,~\cite{jung2021accelerating} uses GPUs at the hardware level to accelerate HE multiplication by 4.05 times in the CKKS scheme. When the kernel method is applied to evaluate the SVM of dimension 784 (single usage: 269$\times$), it can increase acceleration by more than 1,076 times. Most current literature focuses on the hardware acceleration of each HE scheme. However, the kernel method does not compete with each accelerator; instead, it \emph{enhances} the time performance of HE circuits.

\noindent\textbf{Reusability of the Kernel Matrix.} The kernel method enables the reusability of the kernel matrix, which can be employed in other kernel algorithms. The server can store both the dataset $\mathbf{D}$ and the kernel matrix $\mathbf{K}$, eliminating the need to reconstruct inner products of data points. This results in faster computations and efficient resource utilization.

% \noindent\textbf{Parallel Computation of Kernel Matrix.} $\phi$-ZER achieves optimal performance when a sufficiently large number of processors are available for the parallel computation of the kernel function $K(\cdot, \cdot)$. This is because the computation of each element of the kernel matrix can be performed independently. Moreover, since the kernel matrix $\mathbf{K}$ is symmetric, the maximum level of parallelism is attained when $\frac{n(n+1)}{2}$ processors are utilized. 

% \noindent\textbf{Memory Efficient.} Once our optimizer decides to perform either general or kernel method, holding both the dataset $\textbf{D}$ and the kernel matrix $\mathbf{K}$ is unnecessary. Thanks to the symmetric property of the kernel matrix, when $d > \frac{n+1}{2}$, it consumes less memory to store the kernel matrix than the data matrix.

% \input{Sections/04_our_model.tex}
\section{Kernel Trick and Asymptotic Complexities}
This section analyzes the time complexities of general and kernel evaluations for ML/STAT algorithms. We first present a summary of the \emph{theoretical} time complexities in Table~\ref{tab:summary_complexity}, followed by an illustration of how these complexities are derived. Note that the $k$-means and $k$-NN algorithms use a Boolean-based approach, while the remaining algorithms are implemented with an arithmetic HE scheme.

\begin{table}[!htb]
\begin{center}
\caption{Comparison of Time Complexities in ML/STAT Algorithms}
\label{tab:summary_complexity}
\scalebox{0.88}{
\begin{tabular}{ccc}
\toprule
\textbf{Algorithm} & \textbf{General TC} & \textbf{Kernel TC} \\
\midrule
SVM & $O(t n^{2} d)$ & $O(t n^{2})$ \\
PCA & $O(m^{2}+mt_{pow}+nm)$ & $O(nr + rt_{pow} + 3n t_{sqrt})$ \\ 
$k$-NN & $O(10ndl + 6ndl^2 + \Delta_{shared})$ & $O(15nl +\Delta_{shared})$ \\
$k$-means & $O(6tnkdl^2 + tnkdl) $ & $O(11t n^2 kl + 3tn k^2 l)$ \\
\midrule
Total Variance & $O(nd)$ & $O(1)$ \\
Distance & $O(d)$ & $O(1)$ \\
Norm & $O(d+3t_{sqrt})$ & $O(3t_{sqrt})$ \\
Similarity & $O(3d+3t_{sinv}+6t_{sqrt})$ & $O(3t_{sinv}+6t_{sqrt})$ \\
\bottomrule
\end{tabular}
}
\end{center}
\end{table}

\subsection{Arithmetic HE Construction}
\noindent\textbf{Matrix Multiplication and Linear Transformation.} Halevi et al.~\cite{halevi2018faster} demonstrate efficient matrix and matrix-vector multiplication within HE schemes. Specifically, the time complexity for multiplying two $n \times n$ matrices is optimized to $O(n^2)$, while linear transformations are improved to $O(n)$.

\noindent\textbf{Dominant Eigenvalue: Power Iteration.} The power iteration method~\cite{golub2000eigenvalue} recursively multiplies a matrix by an initial vector until convergence. For a square matrix of degree $m$, assuming $t_{pow}$ iterations, the time complexity is $O(m t_{pow})$.

% \noindent\textbf{Square Root and Inverse Operation of Real Numbers.} 
% We use Wilkes's iterative algorithm~\cite{wilkes1951preparation} to approximate the square root operation. It requires $O(3 t_{sqrt})$ multiplications where $t_{sqrt}$ refers to the time until convergence. Furthermore, we use Goldschmidt’s division algorithm to approximate the inverse operation, which requires $O(3 t_{sinv})$ complexity; for every iteration, three scalar multiplications are needed.

\noindent\textbf{Square Root and Inverse Operation of Real Numbers.} 
Wilkes's iterative algorithm~\cite{wilkes1951preparation} approximates the square root operation with a complexity of $O(t_{sqrt})$. Goldschmidt’s division algorithm approximates the inverse operation with a complexity of $O(t_{sinv})$.

\subsubsection{SVM}
% \textcircled{\small{1}} \textbf{SVM.}
\emph{Kernel Trick}. The main computation part of SVM is to find the optimal parameter $\bm{\alpha} = (\alpha_1, \dots, \alpha_n)$ to construct the optimal hyperplane $h^*(\mathbf{x}) = \mathbf{w}^T \mathbf{x} + b$.
For optimization, we use SGD algorithm that iterates over the following gradient update rule: $\alpha_k = \alpha_k + \eta (1 - y_k \sum\limits_{i=1}^{n} \alpha_i y_i \mathbf{x}_i^T \mathbf{x}_j)$.
%
% \begin{equation}
% \label{eq:svm_SGD}
%     \alpha_k = \alpha_k + \eta (1 - y_k \sum\limits_{i=1}^{n} \alpha_i y_i \mathbf{x}_i^T \mathbf{x}_j).
% \end{equation}
%

The kernel trick for SVM is a replacement of $\mathbf{x}_i^T \mathbf{x}_j$ with a linear kernel element $K(\mathbf{x}_i, \mathbf{x}_j)$. Thus, we compute $\alpha_k = \alpha_k + \eta (1 - y_k \sum\limits_{i=1}^{n} \alpha_i y_i K(\mathbf{x}_i, \mathbf{x}_j))$
%
% \begin{equation}
% \label{eq:svm_kernel_SGD}
%     \alpha_k = \alpha_k + \eta (1 - y_k \sum\limits_{i=1}^{n} \alpha_i y_i K(\mathbf{x}_i, \mathbf{x}_j))
% \end{equation}
%
for the kernel evaluation.

\noindent\textbf{(1) General Method.} The gradient update rule for $\alpha_k$ requires $n(d+2) + 1 = O(nd)$ multiplications. For the complete set of $\bm{\alpha}$ and assuming that the worst-case of convergence happens in $t$ iterations, the total time complexity is $O(t n^2d)$.

\noindent\textbf{(2) Kernel Method.} The kernel trick bypasses inner products, which requires $d$ multiplications. Hence, it requires $2n+1 = O(n)$ multiplications for updating $\alpha_k$ ($O(n^2)$ for $\bm{\alpha}$). Therefore, the total time complexity is $O(t n^2)$.

\subsubsection{PCA}
% \textcircled{\small{2}} \textbf{PCA.}
\emph{Kernel Trick}. Suppose $\lambda_1$ is the dominant eigenvalue with the corresponding eigenvector $\mathbf{u}_1$ of the covariance matrix $\bm{\Sigma}$. Then, the eigenpair $(\lambda_1, \mathbf{u}_1)$ satisfies $\bm{\Sigma} \mathbf{u}_1 = \lambda_1 \mathbf{u}_1$. Since $\bm{\Sigma} = \frac{1}{n} \sum\limits_{i=1}^n \mathbf{x}_i \mathbf{x}_i^T$, we can express $\mathbf{u}_1 = \sum\limits_{i=1}^n c_i \mathbf{x}_i$ where $c_i = \frac{\mathbf{x}_i^T \mathbf{u}_1}{n \lambda_1}$. Plugging in the formulae for $\Sigma$ and $\bm{\mu}_1$ and multiplying $\mathbf{x}_k^T$ on both sides yield: 
%
%\begin{equation*}
%\label{eq:pca_kernel_trick1}
$    \sum\limits_{i=1}^n \mathbf{x}_k^T \mathbf{x}_i \sum\limits_{j=1}^n c_j \mathbf{x}_i^T \mathbf{x}_j = n \lambda_1 \sum \limits_{i=1}^n c_i \mathbf{x}_k^T \mathbf{x}_i$
%\end{equation*}
%
for all $\mathbf{x}_k \in \textbf{D}$. By substituting inner products with the corresponding kernel elements $K(\cdot, \cdot)$, we derive the following equation expressed in terms of the kernel elements:
\begin{equation*}
    \sum\limits_{i=1}^n K(\mathbf{x}_k, \mathbf{x}_i) \sum\limits_{j=1}^n c_j K(\mathbf{x}_i, \mathbf{x}_j) = n \lambda_1 \sum \limits_{i=1}^n c_i K(\mathbf{x}_k, \mathbf{x}_i).
\end{equation*}
Further simplification yields $\mathbf{K} \mathbf{c} = (n \lambda_1) \mathbf{c}$, where $\mathbf{c}$ is an eigenvector of the kernel matrix $\mathbf{K}$. Thus, a reduced basis element $\bm{\mu}_i$ is derived by scaling $\mathbf{c}_i$ with $\sqrt{\lambda_i \mathbf{c}_i^T \mathbf{c}_i}$.

\noindent\textbf{Comparison of Complexities.} Both general and kernel methods require the computation of eigenvalues and eigenvectors using the power iteration method for the covariance matrix $\bm{\Sigma}$ and kernel matrix $\mathbf{K}$.

\noindent\textbf{(1) General Method.} This algorithm involves one matrix multiplication, power iteration, and $n$ matrix-vector multiplications. The time complexity is $O(m^2 + m t_{pow} + nm)$, where $m = \text{max}(n, d)$.

\noindent\textbf{(2) Kernel Method.} This algorithm involves power iteration, $r$ normalizations (square root), and $rn$ inner-products. The time complexity is $O(n t_{pow} + 3r t_{sqrt} + nr)$.

\subsubsection{Total Variance}

\emph{Kernel Trick}. The linear kernel trick simplifies the normal variance formula by expressing it in terms of dot products:
{\small\begin{align*}
    TV = \frac{1}{n} \sum\limits_{i=1}^n \lVert \mathbf{x}_i - \bm{\mu} \rVert^2 
    = \frac{1}{n} \sum\limits_{i=1}^n K(\mathbf{x}_i, \mathbf{x}_i) - \frac{1}{n^2} \sum\limits_{i=1}^n \sum\limits_{j=1}^n K(\mathbf{x}_i, \mathbf{x}_j) 
\end{align*}}where the second equality holds by substituting $\bm{\mu} = \frac{1}{n} \sum_{i=1}^n \mathbf{x}_i$.

\noindent\textbf{(1) General Method.} This requires $n$ dot products of deviations; the time complexity is $O(nd)$.

\noindent\textbf{(2) Kernel Method.} The kernel trick for total variance bypasses dot products, resulting in a time complexity of $O(1)$.

\subsubsection{Distance / Norm} 
% \textcircled{\small{4}} \textbf{Distance / Norm.}
\emph{Kernel trick}. We can apply the linear kernel trick on distance $\mathcal{d}(\mathbf{x}_i, \mathbf{x}_j) = \norm{\mathbf{x}_i - \mathbf{x}_j}^2$:
%
% \begin{align*}
% \begin{minipage}{\linewidth}
% \begin{center}
% \resizebox{\linewidth}{!} { 
%     $d(\mathbf{x}_i, \mathbf{x}_j) = \lVert \mathbf{x}_i \rVert^2 - 2 \mathbf{x}_i^T \mathbf{x}_j + \lVert \mathbf{x}_j \rVert^2 = K(\mathbf{x}_i, \mathbf{x}_i) - 2 K(\mathbf{x}_i, \mathbf{x}_j) + K(\mathbf{x}_j, \mathbf{x}_j).$
% }
% \end{center}
% \end{minipage}
% \end{align*}
%
\begin{align*}
\begin{split}
    \mathcal{d}(\mathbf{x}_i, \mathbf{x}_j) &= \lVert \mathbf{x}_i \rVert^2 - 2 \mathbf{x}_i^T \mathbf{x}_j + \lVert \mathbf{x}_j \rVert^2 \\
    &= K(\mathbf{x}_i, \mathbf{x}_i) - 2 K(\mathbf{x}_i, \mathbf{x}_j) + K(\mathbf{x}_j, \mathbf{x}_j).
\end{split}
\end{align*}
\noindent\textbf{(1) General Method.} The general distance formula requires $O(d)$ for evaluating a dot product. Moreover, norm operation entails additional square root operation; hence, $O(d + 3t_{sqrt})$.

\noindent\textbf{(2) Kernel Method.} The kernel method necessitates only addition and subtraction requiring $O(1)$. Likewise, kernelized norm involves the square root operation; thus, $O(3t_{sqrt})$.

\subsubsection{Similarity} 
% \textcircled{\small{5}} \textbf{Similarity.}
\emph{Kernel Trick}. Applying the linear kernel trick to cosine similarity formula, we have 
\[
    sim(\mathbf{x}_i, \mathbf{x}_j) = \frac{\mathbf{x}_i \cdot \mathbf{x}_j}{\lVert \mathbf{x}_i \rVert \lVert \mathbf{x}_j \rVert} 
    = \frac{K(\mathbf{x}_i, \mathbf{x}_j)}{\sqrt{K(\mathbf{x}_i, \mathbf{x}_i)} \sqrt{K(\mathbf{x}_j, \mathbf{x}_j)}}.
\]

\noindent\textbf{(1) General Method.} We compute three dot products, one multiplication, one inverse of a scalar, and two square root operations for computing the general cosine similarity. Hence, it requires $O(3d + 3t_{sinv} + 6t_{sqrt})$.

\noindent\textbf{(2) Kernel Method.} We can bypass the evaluation of the three inner products using the kernel method. Thus, the total time complexity is $O(3t_{sinv} + 6t_{sqrt})$. 

\subsection{Boolean-based HE Construction}
\noindent\textbf{$k$-means and $k$-NN Boolean Construction.} 
Performing $k$-means and $k$-NN in the encrypted domain requires functionalities beyond addition and multiplication, such as comparison operations. In the $k$-means algorithm, it is essential to evaluate the minimum encrypted distance to label data based on proximity to each cluster's encrypted mean. Similarly, $k$-NN necessitates evaluating the maximum encrypted labels given the encrypted distances to all data points. Detailed algorithms for $k$-means and $k$-NN are provided in the Appendix.

\noindent\textbf{Time Complexity for TFHE Circuit Evaluation.} We measure the algorithms' performance by counting the \emph{total number of binary gates} for exact calculation. Boolean gates such as XOR, AND, and OR take twice the time of a single MUX gate~\cite{chillotti2020tfhe}. We assume that the TFHE ciphertext is an encryption of an $l$-bit fixed-point number, with $l/2$ bits for decimals, $(l/2 - 1)$ bits for integers, and $1$ bit for the sign bit. Table~\ref{tab:complexity_tfhe_basic_operations} presents the time complexity of fundamental TFHE operations used throughout the paper.
\begin{table*}[htb!]
\caption{Time complexity of TFHE operations used in the paper}
\label{tab:complexity_tfhe_basic_operations}

\begin{center}
\scalebox{1}{
\begin{tabular}{ccccc ccc}
\toprule

 & \multicolumn{7}{c}{\textbf{TFHE Basic Operation}} \\

 & Comparison & argmin / argmax & Add / Subt. & Multiplication & Absolute Value & Two's Complement & Division \\

\midrule
$\#$(Binary Gate) & $3l$ & $k(k-1)(3l+1)$ & $5l-3$ & $6l^2 + 15l - 6$ & $4l-1$ & $2l - 1$ & $\frac{27}{2} l^2 - \frac{3}{2} l + 1$ \\ 
 
\bottomrule

\end{tabular}}
\end{center}
\end{table*}

% \subsubsection{\texorpdfstring{$k-$means and $k-$NN Boolean Construction}{k-means and k-NN Boolean construction}} 

\subsubsection{\texorpdfstring{$k$-means}{k-means}}

\emph{Kernel Trick}. Let $\mathcal{d}_{ij}$ denote the distance from data point $\mathbf{x}_i$ to the mean of cluster $g_j$. 
{\small\begin{equation*}
    \mathcal{d}_{ij} = \lVert \mathbf{x}_i - \bm{\mu}_j \rVert^2 = \mathbf{x}_i^T \mathbf{x}_i - \frac{1}{n_j} \sum_{\mathbf{x}_a \in g_j} \mathbf{x}_i^T \mathbf{x}_a + \frac{1}{n_j^2} \sum_{\mathbf{x}_a, \mathbf{x}_b \in g_j} \mathbf{x}_a^T \mathbf{x}_b    
\end{equation*}}where $n_j$ is the number of elements in cluster $g_j$. Applying the linear kernel trick, we can compute $\mathcal{d}_{ij}$ using kernel elements:
{\small\begin{equation*}
    \mathcal{d}_{ij} = K(\mathbf{x}_i, \mathbf{x}_i) - \frac{1}{n_j} \sum_{\mathbf{x}_a \in g_j} K(\mathbf{x}_i, \mathbf{x}_a) + \frac{1}{n_j^2} \sum_{\mathbf{x}_a, \mathbf{x}_b \in g_j} K(\mathbf{x}_a, \mathbf{x}_b)
\end{equation*}}

Further simplification yields the objective function:
{\small\begin{equation}
    j^* = \underset{j}{\text{argmin}} \left\{ -n_j \sum_{\mathbf{x}_a \in g_j} K(\mathbf{x}_i, \mathbf{x}_a) + \sum_{\mathbf{x}_a, \mathbf{x}_b \in g_j} K(\mathbf{x}_a, \mathbf{x}_b) \right\}
\label{eq:kernel_kmeans_obj}
\end{equation}}where the common factor $K(\mathbf{x}_i, \mathbf{x}_i)$ and divisions are omitted for efficiency.

\noindent\textbf{(1) General Method.} The $k$-means algorithm involves four steps: computing distances $\mathcal{d}_{ij}$, labeling $\mathbf{x}_i$, forming labeled dataset $\textbf{D}^{(i)}$, and calculating cluster means $\bm{\mu}_j$. The total complexity of $k$-means is asymptotically $O(6tnkdl^2 + tnkdl)$, assuming $t$ iterations until convergence. 

% (See Table~\ref{tab:kmeans_by_submodules} in Appendix~\ref{appendix: kmeans}).

\noindent\textbf{(2) Kernel Method.} The kernel $k$-means algorithm involves four steps: computing the labeled kernel matrix $K^{(j)}$, counting the number of elements $\mathcal{n}_j$ in each cluster, computing distances $\mathcal{d}_{ij}$, and labeling $\mathbf{x}_i$. The total complexity of kernel $k$-means is asymptotically $O(11t n^2 kl + 3tn k^2 l)$, assuming $t$ iterations until convergence. 

\subsubsection{\texorpdfstring{$k$-NN}{k-NN}}

\emph{Kernel Trick}. The linear kernel trick can be applied to the distance $\mathcal{d}_i$ in $k$-NN:
\begin{equation*}
\begin{split}
    \mathcal{d}_i = \lVert \mathbf{x} - \mathbf{x}_i \rVert^2 &= \mathbf{x}^T \mathbf{x} - 2 \mathbf{x}^T \mathbf{x}_i + \mathbf{x}_i^T \mathbf{x}_i \\
    &= K(\mathbf{x}, \mathbf{x}) - 2K(\mathbf{x}, \mathbf{x}_i) + K(\mathbf{x}_i, \mathbf{x}_i).
\end{split}
\end{equation*}
For computational efficiency, we simplify $\mathcal{d}_i$ by removing the common kernel element $K(\mathbf{x}, \mathbf{x})$:
\begin{equation}
    \mathcal{d}_i = -2 K(\mathbf{x}, \mathbf{x}_i) + K(\mathbf{x}_i, \mathbf{x}_i).
\label{eq:kernel_knn}
\end{equation}

\noindent\textbf{Comparison of Complexities.} Both $k$-NN algorithms share the same processes: sorting, counting, and finding the majority label. Specifically, sorting requires $(n-1)^2$ swaps, with each swap involving 4 MUX gates and one comparison circuit, resulting in a complexity of $11(n-1)^2 l$. Counting elements in each class requires $ks$ comparisons and additions, with a complexity of $ks(8l-3)$. Finding the majority index among $s$ labels has a complexity of $s(s-1)(3l+1)$. Thus, the shared complexity is $O(11n^2 l + 8ksl + 3s^2 l)$, denoted as $\Delta_{shared}$.

\noindent\textbf{(1) General Method.} Computing $\mathcal{d}_i = \lVert \mathbf{x} - \mathbf{x}_i \rVert^2$ involves $d$ subtractions, $d$ multiplications, and $d-1$ additions, resulting in a total complexity of $O(10ndl + 6ndl^2)$. Including the shared process, the total complexity is $O(10ndl + 6ndl^2 + \Delta_{shared})$.

\noindent\textbf{(2) Kernel Method.} Using Eq.~(\ref{eq:kernel_knn}), $\mathcal{d}_i$ is computed with one addition and two subtractions per element, totaling $3n$ additions. Including the shared process, the total complexity is $O(15nl + \Delta_{shared})$.

\section{Experiment}
\subsection{Evaluation Metric}
We aim to evaluate the effectiveness of the kernel method in various ML/STAT algorithms using the following metrics.

\noindent\textbf{(1) Kernel Effect in HE.} Let $t_{gen}$ and $t_{ker}$ denote the execution times for the general and kernel methods, respectively. The \emph{kernel effectiveness} or speed-up is evaluated by the ratio:
\begin{equation*}
    EFF = \frac{t_{gen}}{t_{ker}}.
\end{equation*}

\noindent\textbf{(2) Kernel Effect Comparison: Plain and HE.} Let $t_{gen}^{PL}$ and $t_{ker}^{PL}$ represent the execution times for the general and kernel methods in the plain domain. Similarly, let $t_{gen}^{HE}$ and $t_{ker}^{HE}$ represent the execution times in the HE domain. The effectiveness of the kernel methods in both domains can be compared by their respective ratios: $EFF^{PL}$ and $EFF^{HE}$. 

% Moreover, we search for a critical point in each algorithm where the general and kernel approach performance intersects. 

\subsection{Experiment Setting}
\noindent\textbf{Environment.} Experiments were conducted on an Intel Core i7-7700 8-Core 3.60 GHz, 23.4 GiB RAM, running Ubuntu 20.04.5 LTS. We used TFHE version 1.1 for Boolean-based HE and CKKS from OpenFHE~\cite{al2022openfhe} for arithmetic HE.

\noindent\textbf{Dataset and Implementation Strategy.} We used a randomly generated dataset of $n=10$ data points with dimensions in $[n/2, 3n/2]$. The dataset is intentionally small due to the computational time required for logic-gates in TFHE. For example, $k$-means with parameters ($n, d=10$) under the TFHE scheme takes 8,071 seconds. Despite the small dataset, this work aligns with on-going efforts like Google's Transpiler~\cite{gorantala2021general} and its extension HEIR, which use TFHE for practical HE circuit construction. Employing the kernel method can significantly enhance scalability and efficiency.

% \noindent\textbf{Implementation Strategy and Dataset.} $k$-means and $k$-NN were implemented using TFHE, while the rest were implemented using CKKS. We utilized a randomly generated dataset of $n=10$ data points with varying dimensions. Time performance was measured across dimensions ranging from $[n/2, 3n/2]$. The dataset size is intentionally small due to the substantial time required for computations. For instance, SVM with parameters ($n=10, d=10$) takes 2,356 seconds.

\noindent\textbf{Parameter Setting.} Consistent parameters were used for both the general and kernel methods across all algorithms.

\noindent\textbf{(1) TFHE Construction.} TFHE constructions were set to a 110-bit security level\footnote{Initial tests at 128-bit security were revised due to recent attacks demonstrating a lower effective security level.}, with 16-bit message precision.

\noindent\textbf{(2) CKKS Construction.} CKKS constructions used a 128-bit security level and a leveled approach to avoid bootstrapping. Parameters ($N, \Delta, L$) were pre-determined to minimize computation time (see Table~\ref{tab:param_ckks_impl}).

\noindent\textbf{(3) $k$-means and $k$-NN.} For the experiments, $k=3$ and $s=2$ were used for $k$-means, and $k=3$ for $k$-NN.

% parameter setting for ckks implementation
\begin{table}[htb!]
\caption{Parameter setting for CKKS implementation}
\centering
\scalebox{1}{
\begin{tabular}{c cccc}
\toprule
 
 & $\lambda$ & RingDim $N$ & ScalingMod $\Delta$ & MultDepth $L$ \\
 
\midrule

SVM & $128$ & $2^{16}$ & $2^{50}$ & $110$ \\

% \rowcolor{Gray}
% Ridge & $128$ & $2^{16}$ & $2^{50}$ & $54$ \\

\rowcolor{Gray}
PCA & $128$ & $2^{16}$ & $2^{50}$ & $39$ \\

% \rowcolor{Gray}
% LDA & $128$ & $2^{16}$ & $2^{50}$ & $65$ \\

Variance & $128$ & $2^{16}$ & $2^{50}$ & $33$ \\

\rowcolor{Gray}
Distance & $128$ & $2^{16}$ & $2^{50}$ & $33$ \\

Norm & $128$ & $2^{16}$ & $2^{50}$ & $33$ \\

\rowcolor{Gray}
Similarity & $128$ & $2^{16}$ & $2^{50}$ & $50$ \\

\bottomrule
\end{tabular}
}

\label{tab:param_ckks_impl}
\end{table}

\section{Results and Analysis}
\label{section:result}

\subsection{Kernel Effect in the Encrypted Domain} 

% ML algorithms in ctx kernel and nonkernel
% \begin{figure}[htb]
%     \centering
%     \begin{subfigure}[t]{0.23\textwidth}
%         % \centering
%         \includegraphics[width=\textwidth]{figures/mix/mix_svm.eps}
%         \caption{SVM}
%     \end{subfigure}
%     \hfill
%     \begin{subfigure}[t]{0.23\textwidth}
%         % \centering
%         \includegraphics[width=\textwidth]{figures/mix/mix_pca.eps}
%         \caption{PCA}
%     \end{subfigure}
%     \hfill
    
%     \begin{subfigure}[t]{0.23\textwidth}
%         \includegraphics[width=\textwidth]{figures/mix/mix_kmeans.eps}
%         \caption{$k-$means}
%     \end{subfigure}
%     \hfill
%     \begin{subfigure}[t]{0.23\textwidth}
%         \includegraphics[width=\textwidth]{figures/mix/mix_kNN.eps}
%         \caption{$k-$NN}
%     \end{subfigure}
%     \hfill
    
%     \caption{Comparison of execution time between ML's general and kernel methods. The figure indicates that the kernel evaluation exhibits nearly-constant or \emph{dimensionless} property.}
%     \label{fig:gen_ker_cipher}
% \end{figure}

\begin{figure}[htb]
    \centering
    \subfloat[\footnotesize SVM]{\includegraphics[width=0.23\textwidth]{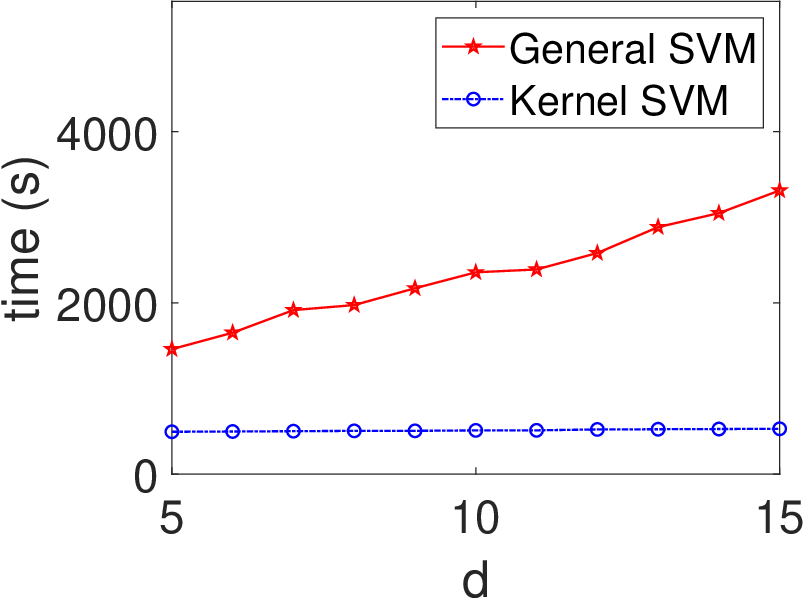}
    \label{fig:mix_svm}}
    \hfill
    \subfloat[\footnotesize PCA]{\includegraphics[width=0.23\textwidth]{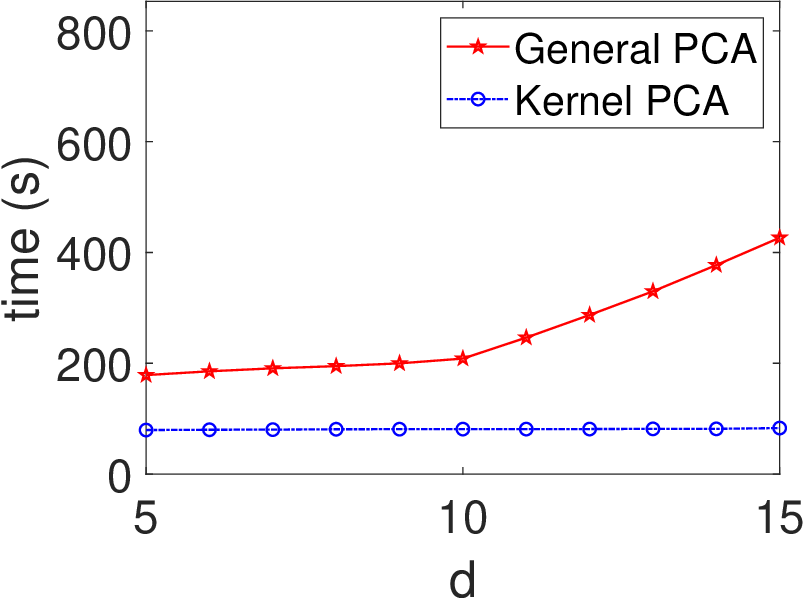}
    \label{fig:mix_pca}}
    \hfill
    
    \subfloat[\footnotesize $k-$means]{\includegraphics[width=0.23\textwidth]{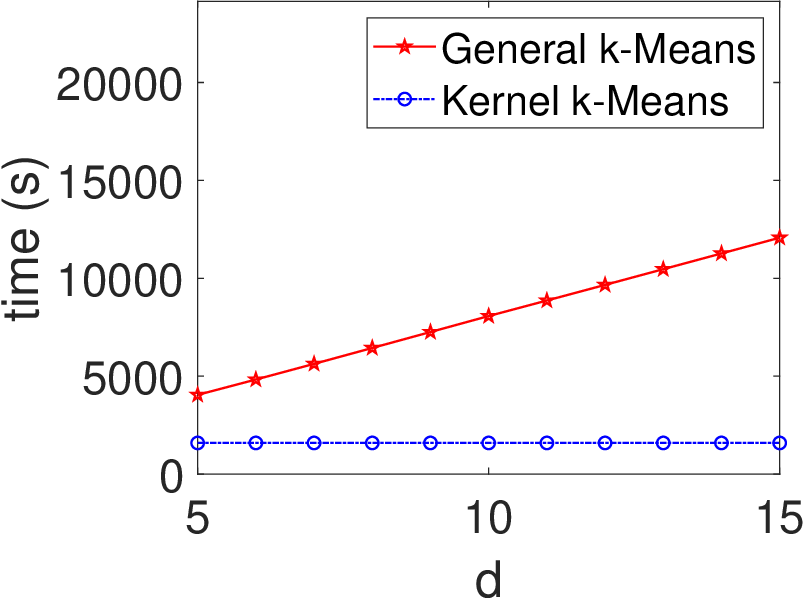}
    \label{fig:mix_kmeans}}
    \hfill
    \subfloat[\footnotesize $k-$NN]{\includegraphics[width=0.23\textwidth]{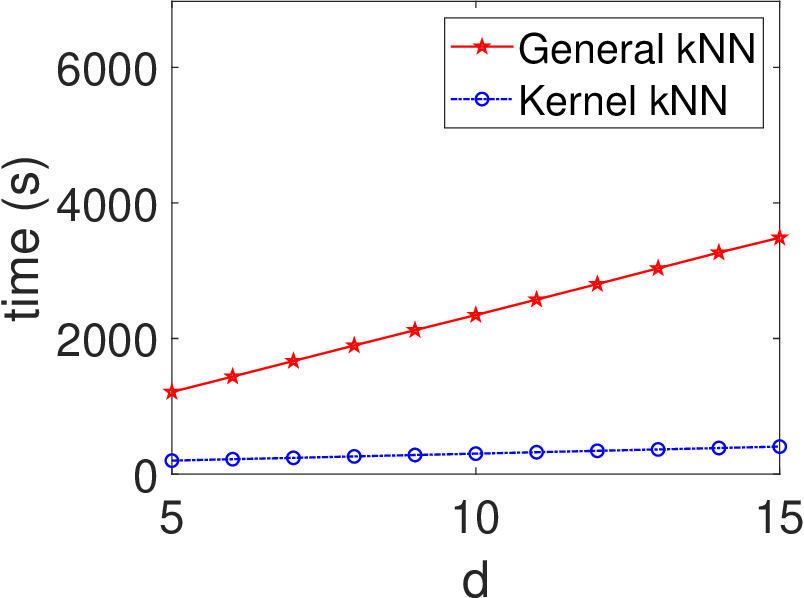}
    \label{fig:mix_knn}}
    \hfill
    
    \caption{Execution Time Comparison Between ML's General and Kernel Methods Showing Kernel's Dimensionless Property (P2).}
    \label{fig:gen_ker_cipher}
\end{figure}

\noindent\textbf{Kernel Method in ML/STAT: Acceleration in HE Schemes.} Fig.~\ref{fig:gen_ker_cipher} demonstrates the stable execution times of kernel methods in ML algorithms within encrypted domains, contrasting with the linear increase observed in general methods. This consistency in both CKKS and TFHE settings aligns with our complexity analysis in Table~\ref{tab:summary_complexity}, improving computational efficiency for algorithms such as SVM, PCA, $k$-means, and $k$-NN. Notably, for general PCA, we observe a linear time increase for $d < n$ and a rapid rise as $O(m^2)$ for $d > n$ due to the quadratic complexity of evaluating the covariance matrix $\bm{\Sigma} = \textbf{D}^T \textbf{D}$.

\begin{figure}[htb]
    \centering
    \subfloat[\footnotesize Total variance]{\includegraphics[width=0.23\textwidth]{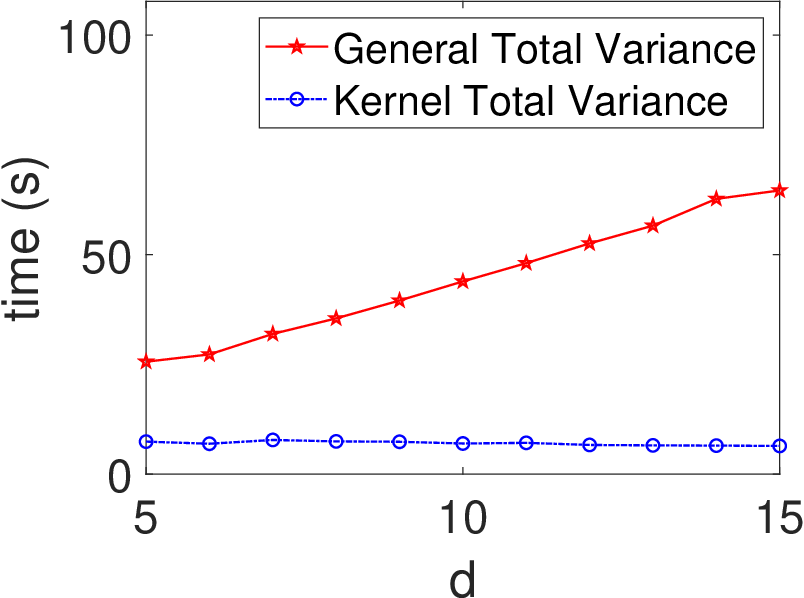}
    \label{fig:mix_tv}}
    \hfill
    \subfloat[\footnotesize Distance]{\includegraphics[width=0.23\textwidth]{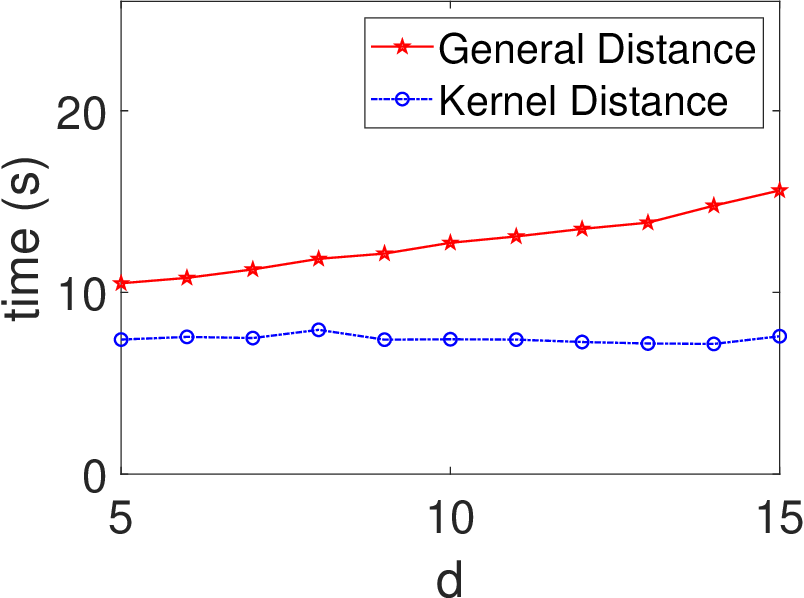}
    \label{fig:mix_distance}}
    \hfill
    
    \subfloat[\footnotesize Norm]{\includegraphics[width=0.23\textwidth]{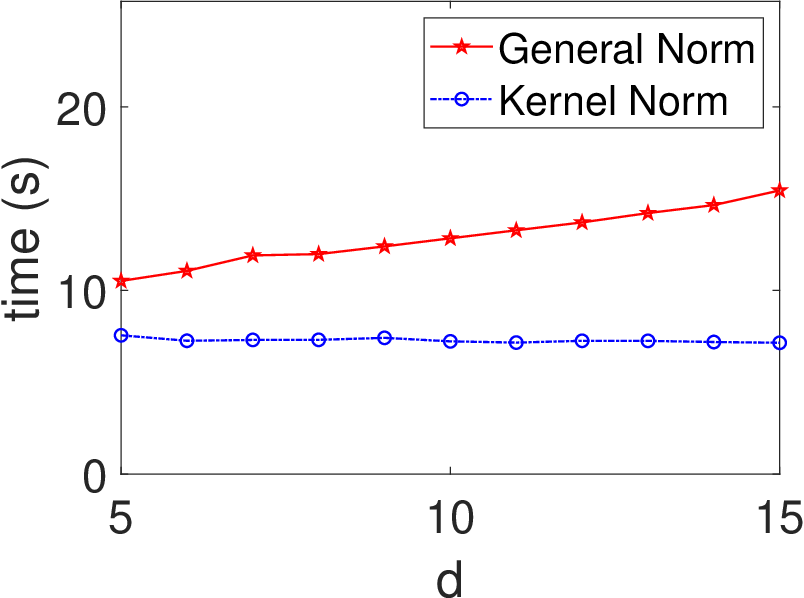}
    \label{fig:mix_norm}}
    \hfill
    \subfloat[\footnotesize Similarity]{\includegraphics[width=0.23\textwidth]{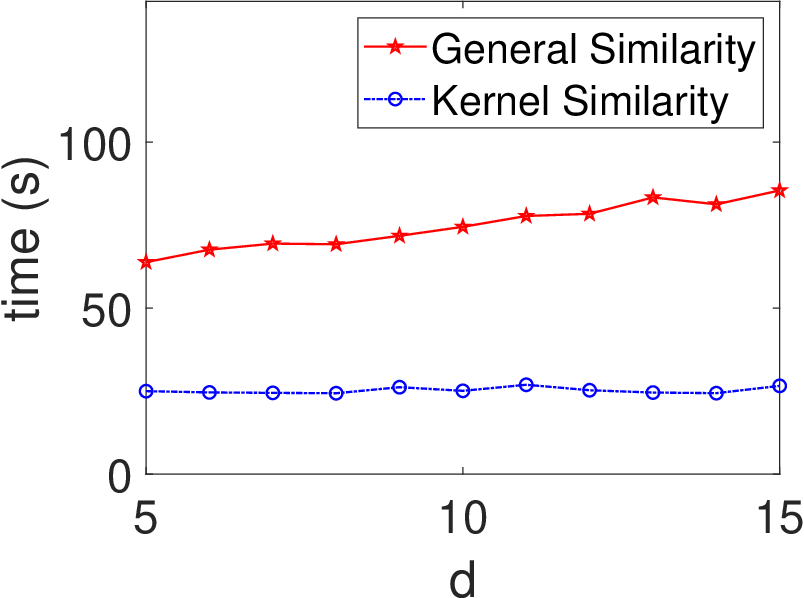}
    \label{fig:mix_sim}}
    \hfill
    
    \caption{Execution Time Comparison Between STAT's General and Kernel Methods Showing Kernel's Outperformance.}
    \label{fig:stats_gen_ker_cipher}
\end{figure}

Fig.~\ref{fig:stats_gen_ker_cipher} shows that the kernel method consistently outperforms general approaches in STAT algorithms within HE schemes. The kernel method maintains stable execution times across various statistical algorithms, unlike the linearly increasing times of general methods as data dimension grows. This aligns with our complexity calculations (Table~\ref{tab:summary_complexity}), highlighting the kernel method's efficiency in encrypted data analysis. The \emph{dimensionless} property of the kernel method makes it a powerful tool for high-dimensional encrypted data analysis, enabling efficient computations in complex settings.

% \noindent\textbf{Critical Point (Kernel Selection Phase).} Our experimental results indicate that for machine learning algorithms such as SVM, PCA, $k-$means, and $k-$NN, there is no critical point in terms of time performance between general and kernel approaches (see Fig.~\ref{fig:gen_ker_cipher}). Specifically, the kernelized circuit consistently outperforms the general circuit, rendering the latter unnecessary. Similarly, all statistical algorithms show better time performance for kernel evaluation in the encrypted domain without any critical point (see Fig.~\ref{fig:stats_gen_ker_cipher}). Based on these observations, our $\phi$-ZER framework (see Fig.~\ref{fig:phizer_framework}) exclusively employs the kernel method for evaluating these algorithms.

\noindent\textbf{Kernelization Time (Pre-Processing Stage).} Kernelization occurs during preprocessing, with kernel elements computed in parallel. Although not included in the main evaluation duration, kernelization time is minimal, constituting less than 0.05 of the total evaluation time. For instance, in SVM, kernel generation accounts for only 0.02 of the total evaluation time (10.82 seconds of 509.91 seconds). Additionally, the reusability of kernel elements in subsequent applications further reduces the need for repeated computations.

\begin{figure*}[htb]
    \centering
    \subfloat[\footnotesize SVM]{\includegraphics[width=0.19\textwidth]{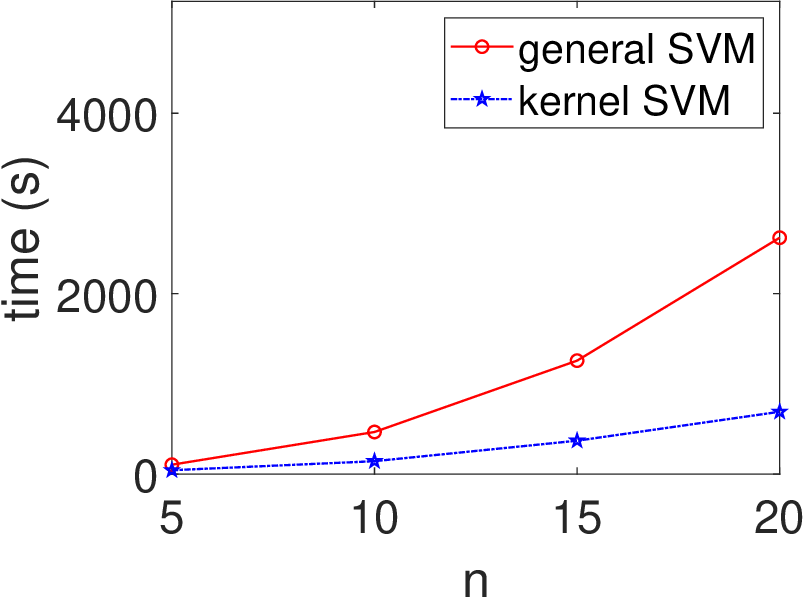}
    \label{fig:svm_by_num}}
    \hfill
    \subfloat[\footnotesize PCA]{\includegraphics[width=0.19\textwidth]{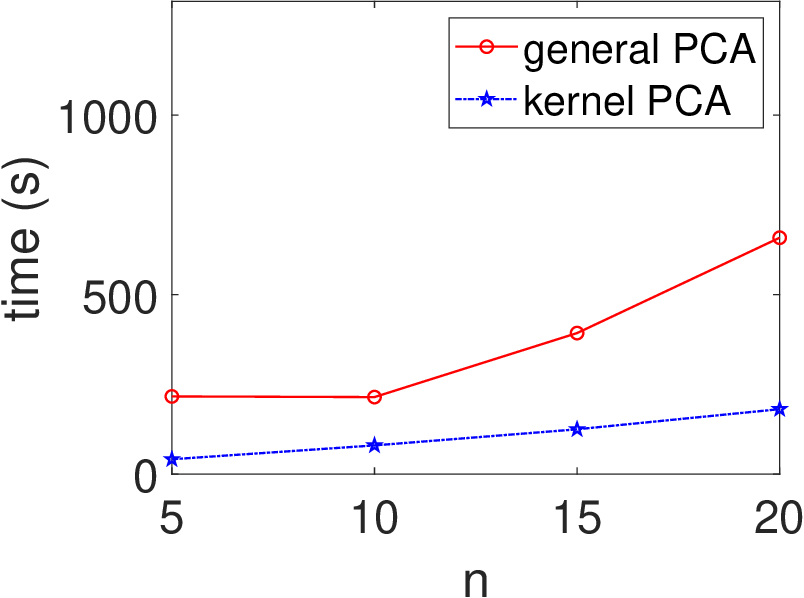}
    \label{fig:pca_by_num}}
    \hfill
    \subfloat[\footnotesize $k-$NN]{\includegraphics[width=0.19\textwidth]{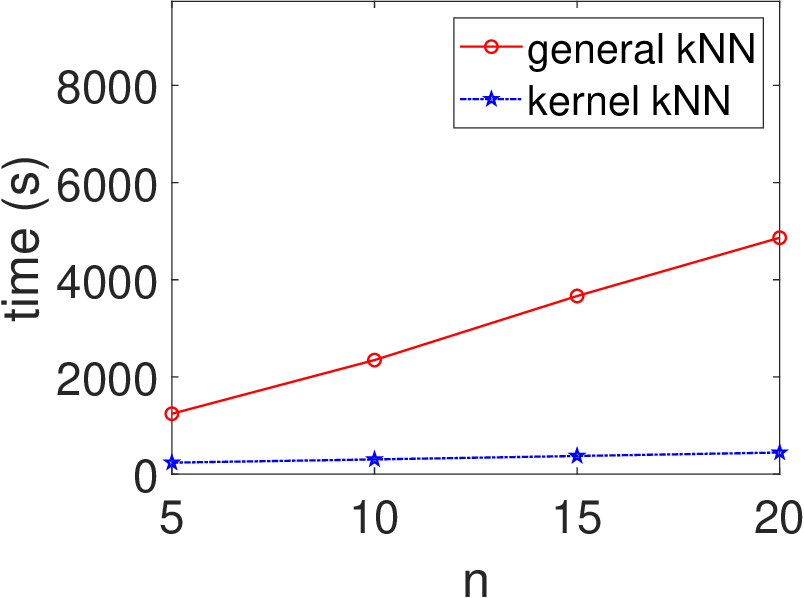}
    \label{fig:knn_by_num}}
    \hfill
    \subfloat[\footnotesize $k-$means]{\includegraphics[width=0.19\textwidth]{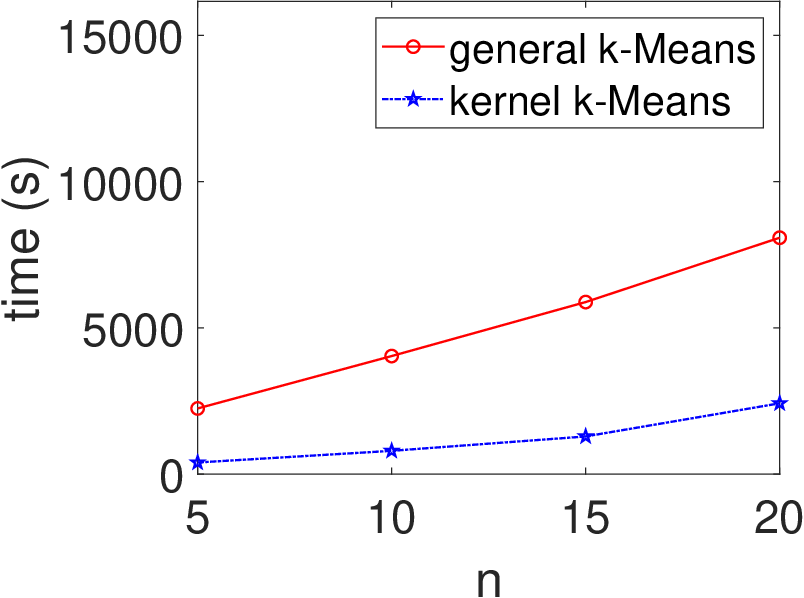}
    \label{fig:kmeans_by_num}}
    \hfill
    \subfloat[\footnotesize Total Variance]{\includegraphics[width=0.19\textwidth]{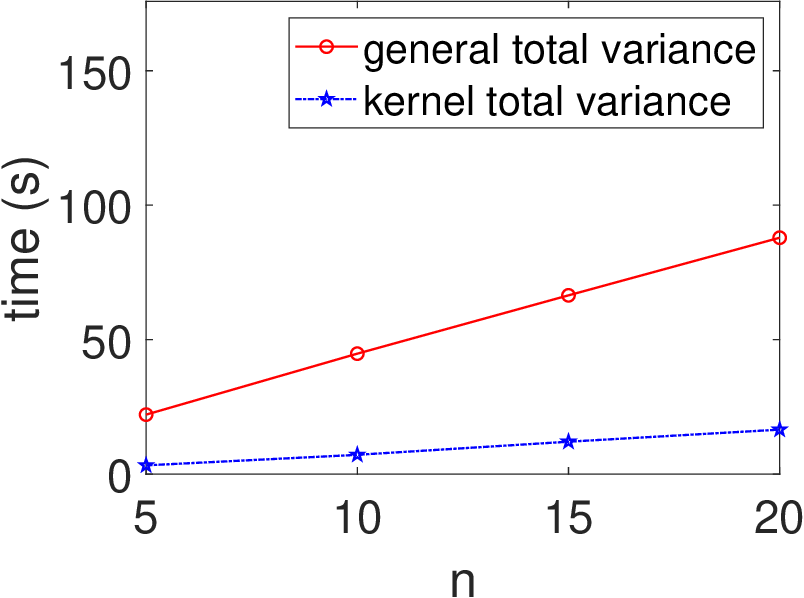}
    \label{fig:total_var_by_num}}
    \hfill
    
    \caption{Comparison of Kernel and General Methods in ML/STAT Algorithms for Different Data Sizes ($n$) at Dimension $d=10$.}
    \label{fig:all_by_num_data}
\end{figure*}

\noindent\textbf{Kernel Method Outperforms with Increasing Data.} We present the execution time of four ML algorithms (SVM, PCA, $k$-NN, $k$-means) and the total variance in the STAT algorithm, with respect to varying numbers of data ($n=5, 10, 15, 20$) at a fixed dimension ($d=10$), as shown in Fig.~\ref{fig:all_by_num_data}. Our results demonstrate that the kernel method consistently outperforms the general method in terms of execution time, even as the number of data increases. This aligns with our complexity table (see Table 3), where general SVM and PCA show a quadratic increase, while $k$-NN, $k$-means, and total variance increase linearly. 

% \noindent\textbf{Time Complexity Analysis of Total Variance.} When examining the total variance of the results presented in Fig~\ref{fig:all_by_num_data}, it can be observed that the kernel total variance increases with the number of data points. Specifically, the kernel total variance increases from 3.23 seconds to 16.55 seconds. However, a comparison to Table~\ref{tab:summary_complexity} reveals that the kernel total variance exhibits constant time complexity of $O(1)$. This is attributed to the $O(n^2)$ increase in the addition operation required for computing the total variance, which involves adding $n^2$ kernel elements: $\sum_{i=1}^n \sum_{j=1}^n K(\mathbf{x}_i, \mathbf{x}_j)$. This highlights the importance of considering the time complexity of addition, especially when multiplication does not significantly contribute to the overall computational complexity, as demonstrated in the case of total variance computation using the kernel method.

\subsection{Kernel Effect Comparison: Plain vs HE}

% ML ptx vs ctx 
\begin{figure}[htb]
    \centering
    \subfloat[\footnotesize SVM]{\includegraphics[width=0.23\textwidth]{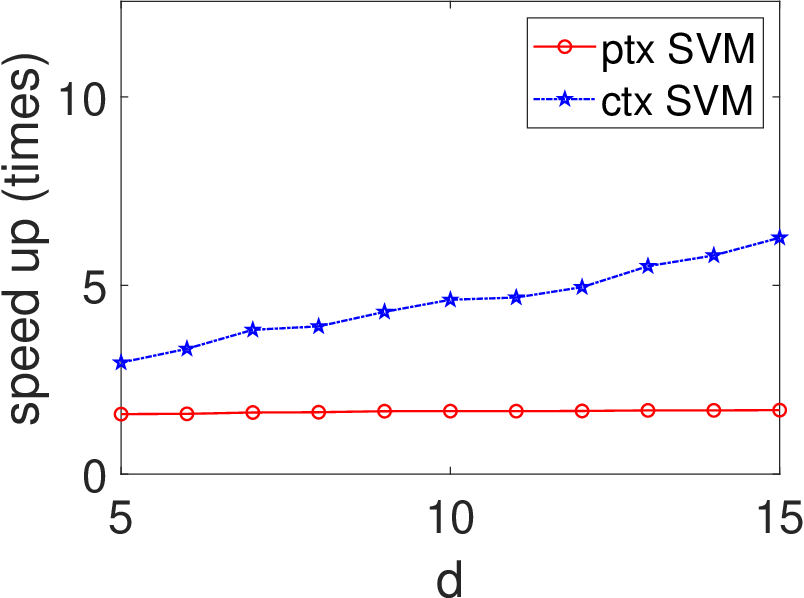}
    \label{fig:svm_ker_effect}}
    \hfill
    \subfloat[\footnotesize PCA]{\includegraphics[width=0.23\textwidth]{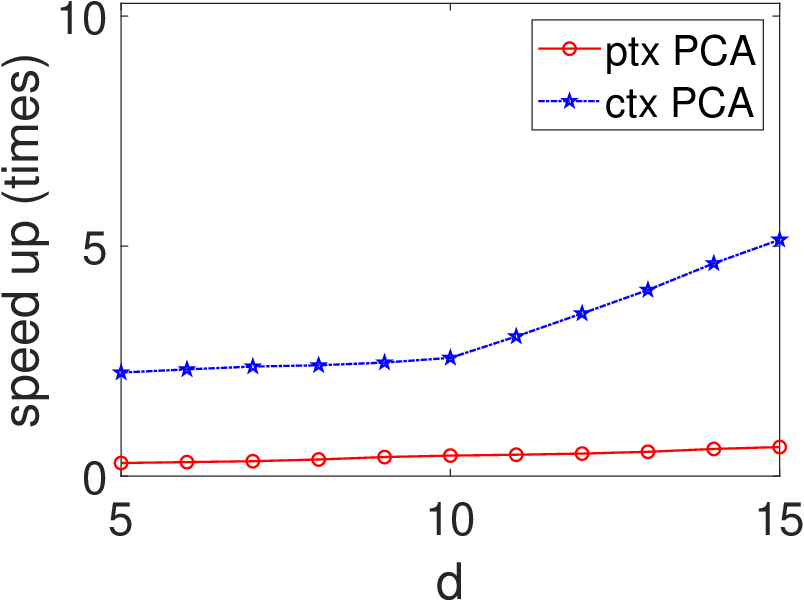}
    \label{fig:pca_ker_effect}}
    \hfill
    
    \subfloat[\footnotesize $k-$means]{\includegraphics[width=0.23\textwidth]{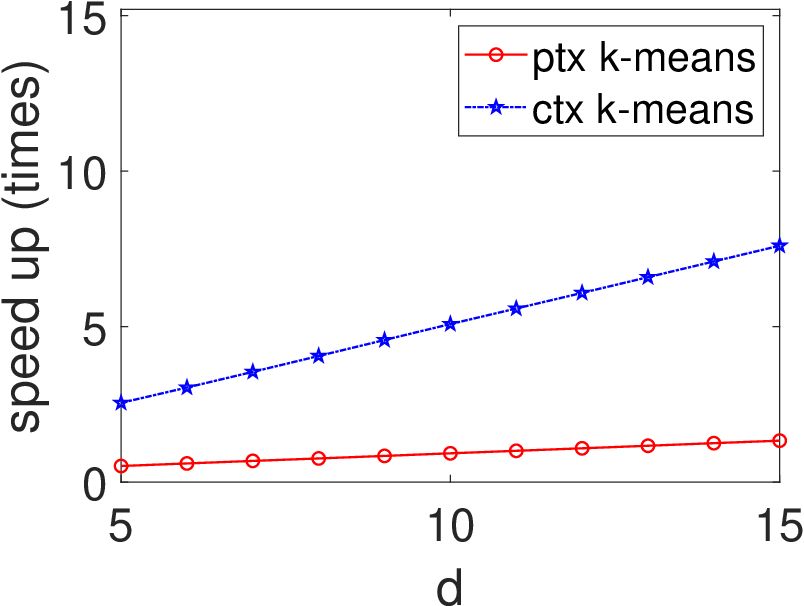}
    \label{fig:kmeans_ker_effect}}
    \hfill
    \subfloat[\footnotesize $k-$NN]{\includegraphics[width=0.23\textwidth]{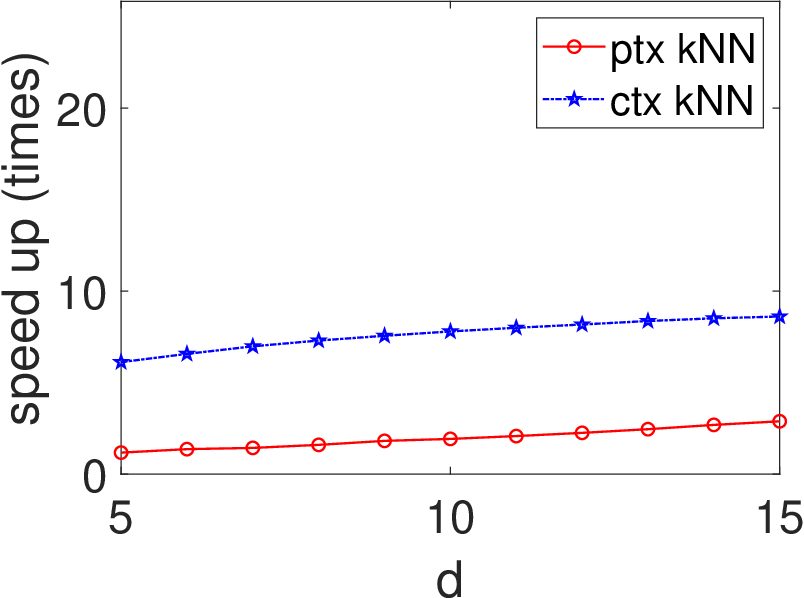}
    \label{fig:knn_ker_effect}}
    \hfill

    \caption{Impact of Kernel Methods in ML Alg. Across Plain and HE Domains.}
    \label{fig:ml_ptx_ctx_ker_effect}
\end{figure}

% STATS ptx vs ctx 
\begin{figure}[htb]
    \centering
    \subfloat[\footnotesize Total variance]{\includegraphics[width=0.23\textwidth]{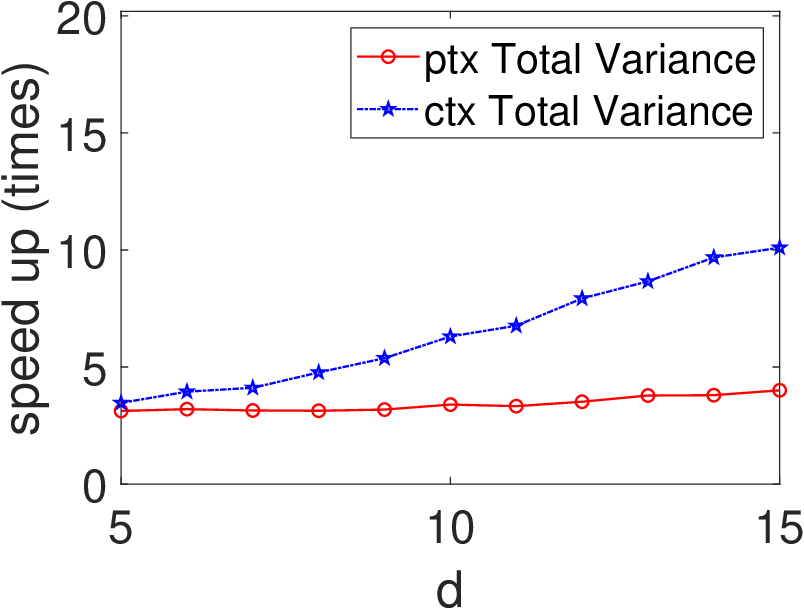}
    \label{fig:tv_ker_effect}}
    \hfill
    \subfloat[\footnotesize Distance]{\includegraphics[width=0.23\textwidth]{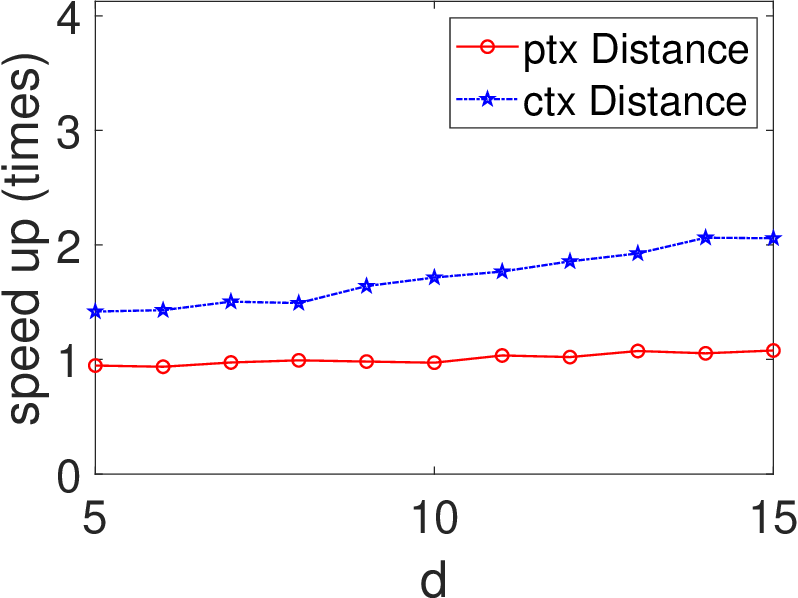}
    \label{fig:dis_ker_effect}}
    \hfill
    
    \subfloat[\footnotesize Norm]{\includegraphics[width=0.23\textwidth]{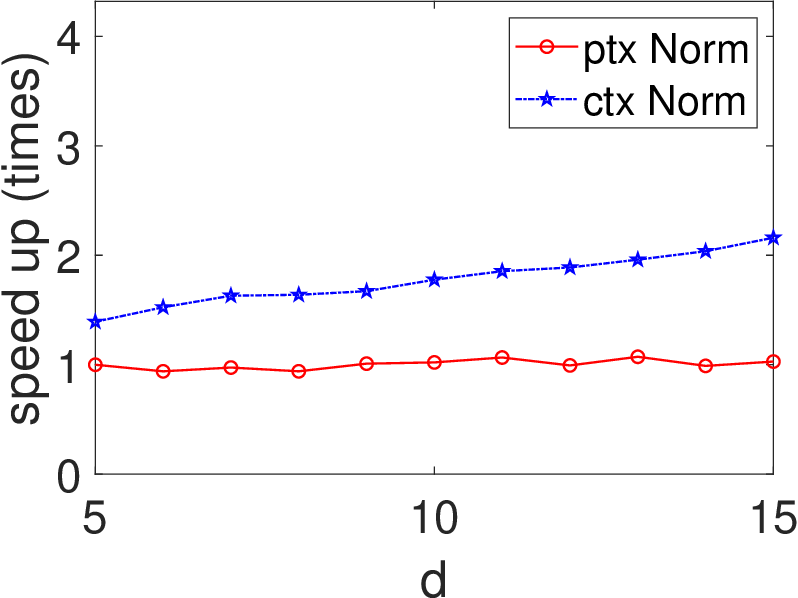}
    \label{fig:norm_ker_effect}}
    \hfill
    \subfloat[\footnotesize Similarity]{\includegraphics[width=0.23\textwidth]{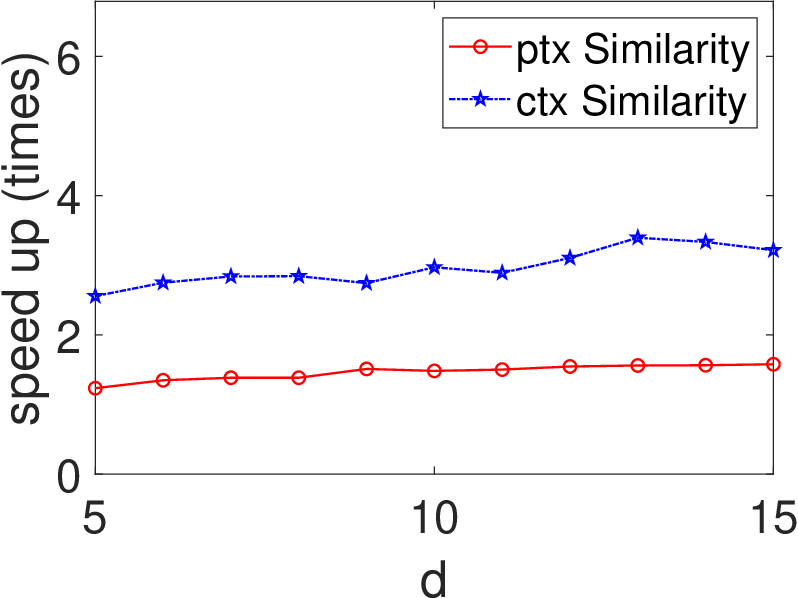}
    \label{fig:sim_ker_effect}}
    \hfill

    \caption{Impact of Kernel Method on STAT Alg. in Plain and HE Domains.}
    \label{fig:stats_ptx_ctx_ker_effect}
\end{figure}

\noindent\textbf{Significant Kernel Effect in the HE Domain.} Our experiments demonstrate a more substantial kernel effect in the HE domain compared to the plain domain for both ML and STAT algorithms. This effect is consistently observed across algorithms (see Fig.~\ref{fig:ml_ptx_ctx_ker_effect} and Fig.~\ref{fig:stats_ptx_ctx_ker_effect}). For example, the kernel effect in SVM within HE amplifies performance by 2.95-6.26 times, compared to 1.58-1.69 times in the plain domain, across dimensions 5-15. This enhanced performance is due to the kernel method's ability to reduce heavy multiplicative operations, a significant advantage in HE where multiplication is more time-consuming than addition (see Section~\ref{sec:ker_effect}).

\section{Conclusion}

This paper introduces the kernel method as an effective optimizer for homomorphic circuits in ML/STAT applications, applicable across various HE schemes. We systematically analyze the kernel optimization and demonstrate its effectiveness through complexity analysis and experiments. Our results show significant performance improvements in the HE domain, highlighting the potential for widespread use in secure ML/STAT applications.

% \noindent \textbf{Reproducibility.} Our source code is available at \url{https://github.com/PrivStatBool/KernelHE}.

% \clearpage

% --- Bibliography

% --- Appendix 
\appendix
% \section*{Appendix}

\section{ML / STAT}

\section{\texorpdfstring{$k-$means and $k-$NN: Boolean Construction}{k-means and k-NN Boolean-based HE Construction}}
% \section{$k-$means and $k-$NN: Boolean Construction}
\label{appendix: boolean_based_construction}

\subsection{\texorpdfstring{Boolean-based HE Construction: $k$-means}{k-means}} 
\label{appendix: boolean_kmeans}

\noindent\textbf{(1) General Method.} Constructing the $k$-means circuit in the encrypted domain presents a primary challenge: the iterative update of encrypted labels $\bm{\mathcal{l}}$ in each iteration.
\begin{equation}
\label{eq:kmeans_argmin}
    l_i = \underset{j}{\text{argmin}} \{ \mathcal{d}_{ij} \}_{j=1}^k.
\end{equation}
This involves two key tasks: 1) labeling data $x_i$ using Eq.~(\ref{eq:kmeans_argmin}), and 2) computing the average value for each cluster $g_i$ based on the encrypted labels.

%% k-means // overall
% \begin{algorithm}[!htb]

% \DontPrintSemicolon
%   \KwInput{data matrix $\textbf{D}$, cluster number $k$, iteration number $t$}
%   \KwOutput{$k$ centroids $\mathbf{M}^t = (\bm{\mu}_1^t, \dots, \bm{\mu}_k^t$) at time $t$}

%   $r \leftarrow 0$

%   $\bm{\mu}_j^r \overset{\$}{\leftarrow}  \chi^d$, $j=1,\dots,k$ \tcp*{input space $\chi$}

%   \For{$r < t$}
%   {

%     \For{$i=1,\dots,n$}
%     {

%     $\mathcal{d}_j \leftarrow \lVert \mathbf{x}_i - \bm{\mu}_j^r \rVert^2, j=1,\dots,k$ 

%     \tcc{Cluster Assignment}

%     $\bm{\mathcal{l}}_i \leftarrow$ TFHE.argmin$_{i}$ $\{ \mathcal{d}_i \}_{i=1,\dots,k}$
    
%     }

%     $\bm{\mathcal{L}} \leftarrow (\bm{\mathcal{l}}_1, \dots, \bm{\mathcal{l}}_n)^T$ \tcp*{labels of $\mathbf{x}_i$}

%     % \tcc{Update Centroid}
    
%     $\mathbf{M}^r \leftarrow$ TFHE.ClusterMean($\textbf{D}, \bm{\mathcal{L}}$)

%     $r \leftarrow r+1$

%   }
% \caption{TFHE.$k$-means($\textbf{D}$, $k$, $t$)}
% \label{alg:kmeans_main}
% \end{algorithm}

\noindent\textbf{Issue 1: Labeling Data.} We solve Eq.~(\ref{eq:kmeans_argmin}) using the TFHE.LEQ$(\mathsf{ct_1}, \mathsf{ct_2})$ comparison operation. This operation returns $\mathsf{Enc}(1)$ if $\mathsf{ct_1}$ is less than or equal to $\mathsf{ct_2}$, and $\mathsf{Enc}(0)$ otherwise. Algorithm~\ref{alg:kmeans_argmin} determines the index of the minimum value from a set of encrypted ciphertexts $\{ \mathcal{d}_i \}_{i=1}^k$. The output is a binary ciphertext $\bm{\mathcal{l}}^* = (\mathcal{l}_1, \dots, \mathcal{l}_k)$, where a non-zero $\mathcal{l}_j$ indicates the label corresponding to $\mathbf{x}_i$.
%
%% k-means // arg min
\begin{algorithm}[!htb]
\DontPrintSemicolon
  % \KwInput{a set of indexed \textsf{ct} $\{ \mathcal{d}_i \}_{i=1,\dots,k}$}
  % \KwOutput{indexed binary \textsf{ct} $\bm{\mathcal{l}}^* = (\mathcal{l}_1, \dots, \mathcal{l}_k)$}

  \For{ $i=1,\dots,k$}
  {
    $\mathcal{l}_i \leftarrow \mathsf{Enc}(1)$

    \For{$j=1,\dots,k$}
    {
        $\mathcal{t} \leftarrow$ $\underset{j \neq i}{\text{TFHE.LEQ}(\mathcal{d}_i, \mathcal{d}_j)}$

        $\mathcal{l}_i \leftarrow$ TFHE.AND($\mathcal{l}_i, t$)

    }
  }
  $\bm{\mathcal{l}}^* \leftarrow (\mathcal{l}_1, \dots, \mathcal{l}_k)$

\caption{TFHE.argmin$_{i}$ $\{ \mathcal{d}_i \}_{i=1,\dots,k}$ }
\label{alg:kmeans_argmin}
\end{algorithm}

% The time complexity of Algorithm~\ref{alg:kmeans_argmin} is $k(k-1)(3l + 1)$, as it compares each distance $\mathcal{d}_i$ with every other $\mathcal{d}_j$.

\noindent\textbf{Issue 2: Computing Average Value.} Identifying the data \( \mathbf{x}_i \) in a specific group \( g_j \) using encrypted labels \( \bm{\mathcal{l}}_i \) is challenging. We address this by extracting cluster data \( \mathbf{D}^{(j)} \), setting non-belonging data to zero through an AND operation between each label \( l_{ij} \) and its respective data \( \mathbf{x}_i \) (see Algorithm~\ref{alg:kmeans_cluster}).

%% k-means // get cluster data
\begin{algorithm}[!htb]
\DontPrintSemicolon
  % \KwInput{data matrix $\mathbf{D}$, label $\bm{\mathcal{L}}$, cluster index $i$}
  % \KwOutput{$\mathbf{D}^{*} = \{\mathbf{x}_j \in \mathbf{D} \, \vert \, \mathbf{x}_j = \mathsf{Enc}(0)  \text{ for } \mathbf{x}_j \not\in g_i \}$ }
  
  \tcc{$\mathbf{x}_j \leftarrow \mathbf{x}_j \text{ or } \mathbf{x}_j \leftarrow \mathsf{Enc}(0) \text{ by label } \mathcal{l}_{ji}$}

  \ForEach{$\mathbf{x}_j = (x_{j1}, \dots, x_{jl}) \in \mathbf{D}$}{
    
    $x_{jk} \leftarrow $ TFHE.AND($\mathcal{l}_{ji}, x_{jk}), k=1,\dots,l$
    
  }
  
\caption{TFHE.GetClusterData($\mathbf{D}$, $\mathcal{L}$)}
\label{alg:kmeans_cluster}
\end{algorithm}

With $\mathbf{D}^{(j)}$ identified, the mean $\bm{\mu}_j$ for each cluster is computed by summing all elements in $\mathbf{D}^{(j)}$ and dividing by the number of elements in the class $g_j$. The number of elements in each class is obtained by summing the encrypted label $\mathcal{l}_{ji}$ for all $\mathbf{x}_i \in \mathbf{D}$ (see Algorithm~\ref{alg:kmeans_cluster_mean}).

%% k-means // cluster mean
\begin{algorithm}[!htb]
\DontPrintSemicolon
  % \KwInput{data matrix $\textbf{D}$, encrypted label $\bm{\mathcal{L}}$}
  % \KwOutput{cluster mean $\textbf{M} = (\bm{\mu}_1, \dots, \bm{\mu}_k)$}

  \For{$i=1,\dots,k$}
  {
  $\mathbf{D}^{(i)} \leftarrow$ TFHE.GetClusterData($\textbf{D}, \bm{\mathcal{L}}$, $i$)

    \For{$\mathbf{x}_j ' \in \mathbf{D}^{(i)}$}
    {
        % \tcc{ use TFHE Add/Div. Bool. circuits }
    
        $\mathcal{n}_i \leftarrow \sum\limits_{i=1}^{n}{\mathcal{l}_{i1}}$, and $\bm{\mu}_i \leftarrow \frac{1}{\mathcal{n}_i} \sum\limits_{j=1}^n \mathbf{x}_j'$
    }

  }
  % $\textbf{M} \leftarrow (\bm{\mu}_1, \dots, \bm{\mu}_k)$

\caption{TFHE.ClusterMean($\textbf{D}$, $\bm{\mathcal{L}}$)}
\label{alg:kmeans_cluster_mean}
\end{algorithm}
%

% Note that we need to compute $k$ times the addition of the whole data $\textbf{D}$ for taking the average of each cluster since the labels for $\mathbf{x}_i$ are encrypted, whereas it requires only one total sum of the data $\mathbf{x}_i \in \textbf{D}$ in the non-encrypted domain.

%--------------------
%------------ kernel k-means
\begin{algorithm}[!htb]
\DontPrintSemicolon
  % \KwInput{data matrix $\textbf{D}$, cluster number $k$, iteration number $t$}
  % \KwOutput{encrypted label matrix $\bm{\mathcal{L}}$}
  
  \tcc{parallel compute linear kernel}
  
  % $\textbf{K} = \{ K(\mathbf{x}_i, \mathbf{x}_j) \}_{i, j = 1, \dots, n} $ 

  $\bm{\mathcal{L}} \leftarrow \{ \bm{\mathcal{l}}_i \vert \bm{\mathcal{l}}_i = \mathsf{Enc}(a \overset{\$}{\leftarrow} \{1,\dots,k\} ), i=1,\dots,n \}$

  % \For{$r < t$}{
    \Repeat{$t$ times}{

      \For{$i=1,\dots,n$}{

          \For{$j=1,\dots,k$}{

          $\textbf{K}^{(j)} \leftarrow $ TFHE.GetClusterData($\textbf{K}, \bm{\mathcal{L}}_i$)

          $\mathcal{n}_j \leftarrow \sum\limits_{s=1}^{n} l_{sj}$
          
          $\mathcal{p} \leftarrow \sum\limits_{a=1}^{n}\sum\limits_{b=1}^{n} \textbf{K}^{(j)}(\mathbf{x}_a, \mathbf{x}_b)$

          $\mathcal{q} \leftarrow -2\mathcal{n}_j \sum\limits_{a=1}^{n} \textbf{K}^{(j)}(\mathbf{x}_a, \mathbf{x}_b)$

          $\mathcal{d}_j \leftarrow \mathcal{p} + \mathcal{q}$
    
          }

          $\bm{\mathcal{l}}_i \leftarrow $ TFHE.argmin$_{j} \{\mathcal{d}_j \}_{j=1,\dots,k}$
        
      }
      % $\bm{\mathcal{L}} \leftarrow (\bm{\mathcal{l}}_1, \dots, \bm{\mathcal{l}}_n)$
  }

\caption{TFHE.Kernel$k$-means($\textbf{D}$, $k$, $t$)}
\label{alg:kernel_kmeans}
\end{algorithm}

\noindent\textbf{(2) Kernel Method.} Kernel $k$-means reduces to solving Eq.~(\ref{eq:kernel_kmeans_obj}). Directly solving partial sums from the kernel matrix is infeasible due to encrypted labels. Therefore, we proceed as follows (see Algorithm~\ref{alg:kernel_kmeans} for the complete kernel evaluation):
\begin{itemize}[leftmargin=0.35cm]
    \item Obtain cluster-specific kernel $K^{(j)}$, where \( K^{(j)}(\mathbf{x}_a, \mathbf{x}_b) \) equals \( K(\mathbf{x}_a, \mathbf{x}_b) \) if both \( \mathbf{x}_a, \mathbf{x}_b \in g_j \); otherwise, it is zero.
    \item Compute the partial sums from Eq.~(\ref{eq:kernel_kmeans_obj}) using the cluster-specific kernel $K^{(j)}$.
\end{itemize}
\noindent\textbf{Proc. 1: Cluster-specific Kernel.} We obtain $K^{(j)}$ using $\mathcal{l}^*_{ij}$ and AND gates:
\begin{align*}
\mathcal{t} & \leftarrow \text{TFHE.AND}(\mathcal{l}^*_{aj}, \mathcal{l}_{bj}), \\
K^{(j)}(\mathbf{x}_a, \mathbf{x}_b)[s] & \leftarrow \text{TFHE.AND}(\mathcal{t}, K(\mathbf{x}_a, \mathbf{x}_b)[s])
\end{align*}
where \( \mathcal{t} \) is \( \mathsf{Enc}(1) \) if both \( \mathbf{x}_a \) and \( \mathbf{x}_b \) belong to \( g_j \). 
% (The bracket notation $[\cdot]$ refers to a bit position of a scalar.)

\noindent\textbf{Proc. 2: Partial Sum Evaluation.} We compute over $n \geq n_j$ elements in $K^{(j)}$:
\begin{align*}
\sum_{\mathbf{x}_a \in g_j} K(\mathbf{x}_i, \mathbf{x}_a) &= \sum_{a=1}^n K^{(j)}(\mathbf{x}_i, \mathbf{x}_a) \\
\sum_{\mathbf{x}_a, \mathbf{x}_b \in g_j} K(\mathbf{x}_a, \mathbf{x}_b) &= \sum_{b=1}^n \sum_{a=1}^n K^{(j)}(\mathbf{x}_a, \mathbf{x}_b)
\end{align*}

%--------------------
%--------------------
%--------------------

\subsection{\texorpdfstring{Boolean-based HE construction: $k-$NN}{k-NN}} 
% \subsection{Boolean-based HE construction: $k-$NN} 
\label{appendix: knn_boolean}

\noindent\textbf{(1) General Method:} The $k$-NN algorithm involves 1) sorting, 2) counting, and 3) finding the majority label among encrypted data. We address the issues as follows.

\noindent\textbf{Issue 1: Sorting.} We use the TFHE.minMax function, detailed in Algorithm~\ref{alg:min_max}, to arrange a pair of ciphertexts \( \mathsf{ct_1}, \mathsf{ct_2} \) in ascending order using the TFHE.LEQ operation. By using \( \mathcal{t} \) and its negation \( \sim \mathcal{t} \) as selectors in a MUX gate, we obtain the ordered pair \( (\mathsf{ct}_{\text{min}}, \mathsf{ct}_{\text{max}}) \). 
% This process is similarly applied to the corresponding encrypted labels \( \mathcal{l}_1 \) and \( \mathcal{l}_2 \). 

%% k-NN overall
% \begin{algorithm}[!htb]
% \DontPrintSemicolon
%   \KwInput{data matrix $\textbf{D} = \{ (\textbf{x}_i, y_i)\}_{i=1}^{n}$, nearest neighbor number $k$, test data $\mathbf{x}$}
%   \KwOutput{predicted class $\hat{y} \in \mathcal{G} =\{g_1, \dots, g_s\}$}

%   \tcc{calculate distance from $\mathbf{x}$ to $\mathbf{x}_i$}
  
%   $\mathcal{d}_i \leftarrow \lVert \mathbf{x} - \mathbf{x}_i \rVert^2, i=1,\dots,n$

%   \tcc{label $y_i^{*}$ coressponds to sorted $\mathcal{d}_i^*$}

%   $(\mathcal{d}_i^*, y_i^{*})_{i=1,\dots,n} \leftarrow $ TFHE.BubbleSort($\bm{\mathcal{d}}, \bm{y}$)

%   \tcc{find majority class for $\mathbf{x}$}

%   $(\mathcal{n}_1, \dots, \mathcal{n}_s) \leftarrow$ TFHE.CountClass($(\mathcal{y}_i)_{i=1,\dots,k}^*$)

%   $\hat{y} \leftarrow$ TFHE.argmax$_j \{ \mathcal{n}_j \}_{j=1, \dots, s}$
  
% \caption{TFHE.$k$-NN($\textbf{D}$, $k$, $\mathbf{x}$)}
% \label{alg:knn_all}
% \end{algorithm}

%% knn / minMax
\begin{algorithm}[!htb]
\DontPrintSemicolon
  % \KwInput{$\mathcal{c}_1, \mathcal{c}_2$, and their corresponding labels $\mathcal{l}_1, \mathcal{l}_2$}
  % \KwOutput{increasing order ($\mathcal{c}^{min}, \mathcal{l}^{min}), (\mathcal{c}^{max}, \mathcal{l}^{max})$)}

  $\mathcal{t} \leftarrow $ TFHE.LEQ($\mathsf{ct}_1, \mathsf{ct}_2$),   $\sim \mathcal{t} \leftarrow $ TFHE.NOT($\mathcal{t}$)

  % \tcc{length $r$ binary form $\mathsf{ct}_i$: $(\mathsf{ct}_{i1}, \dots, \mathsf{ct}_{ir})$}

  \For{$j=1,\dots,r$}
  {

      $\mathsf{ct}_j^{min} \leftarrow$ TFHE.MUX($\mathcal{t}$, $\mathsf{ct}_{1j}$, $\mathsf{ct}_{2j}$)
    
      $\mathsf{ct}_j^{max} \leftarrow$ TFHE.MUX($\sim \mathcal{t}$, $\mathsf{ct}_{1j}$, $\mathsf{ct}_{2j}$)
    
      $\mathcal{l}_j^{min} \leftarrow$ TFHE.MUX($\mathcal{t}$, $\mathcal{l}_{1j}$, $\mathcal{l}_{2j}$)
    
      $\mathcal{l}_j^{max} \leftarrow$ TFHE.MUX($\sim \mathcal{t}$, $\mathcal{l}_{1j}$, $\mathcal{l}_{2j}$)
    
  }
  
\caption{TFHE.minMax($(\mathsf{ct}_1, \mathcal{l}_1), (\mathsf{ct}_2, \mathcal{l}_2)$)}
\label{alg:min_max}
\end{algorithm}

By employing the TFHE.minMax function, Algorithm~\ref{alg:bubble_sort} executes the Bubblesort algorithm, yielding a sorted sequence \( \{d_i^* \}_{i=1}^n \) in ascending order, with their respective labels \( \mathcal{l}_i^* \).

%
%% knn / bubblesort
\begin{algorithm}[!htb]
\DontPrintSemicolon
  % \KwInput{$\bm{\mathcal{b}} = (\mathcal{b}_1, \dots, \mathcal{b}_n)$ and its label $\bm{\mathcal{l}} = (\mathcal{l}_1, \dots, \mathcal{l}_n)$}
  % \KwOutput{sorted $\bm{\mathcal{b}}^*, \bm{\mathcal{l}}^*$}

  $\mathcal{a}_j = (\mathcal{d}_j, \mathcal{l}_j)$

  \Repeat{$n-1$ times}
  {

      \For{$j=1,\dots,n-1$}
      {
    
        $(\mathcal{a}_j, \mathcal{a}_{j+1}) \leftarrow $ TFHE.minMax($\mathcal{a}_j, \mathcal{a}_{j+1}$)
    
      }
    
  }

\caption{TFHE.BubbleSort($\{\mathcal{d}_i \}_{i=1}^n, \bm{\mathcal{l}}$)}
\label{alg:bubble_sort}
\end{algorithm}

\noindent\textbf{Issue 2: Counting.} From the sorted distances \( \{d_i^* \}_{i=1}^n \) and respective labels \( \{\mathcal{l}_i^*\}_{i=1}^n \), we count the number of elements in each class \( g_j \) among the $k$ nearest neighbors. We repeatedly use the TFHE.EQ comparison, which outputs $\mathsf{Enc}(1)$ if two input ciphertexts are equal (see Algorithm~\ref{alg:count_class}). 
% The TFHE.CountClass function scans through the labels \( \{ \mathcal{l}^*_i \}_{i=1}^k \), performing the matching operation TFHE.EQ within the range \( \mathcal{l}^*_i \in \{1, \dots, s\} \). When a match occurs, $\mathsf{Enc}(1)$ is added.

% \vspace{0.1cm}

%% k-means // elts for each class
\begin{algorithm}[!htb]
\DontPrintSemicolon
  % \KwInput{label $\{\mathcal{y}_i\}_{i=1,\dots,k}$, number of class $s$}
  % \KwOutput{number of elements in each class $\bm{\mathcal{n}}$}

  \For{$i=1,\dots,s$}
  {
    $\mathcal{n}_i \leftarrow \mathsf{Enc}(0)$

    \For{$j=1,\dots,k$}
    {
        $\mathcal{t} \leftarrow$ TFHE.EQ($\mathsf{Enc}(i), \mathcal{l}^*_j$)

        $\mathcal{n}_i \leftarrow \mathcal{n}_i + \mathcal{t}$
    }
  }
  % $\bm{\mathcal{n}} \leftarrow (\mathcal{n}_1, \dots, \mathcal{n}_s)$

\caption{TFHE.CountClass($\{\mathcal{l}^*_i\}_{i=1,\dots,k}, s$)}
\label{alg:count_class}
\end{algorithm}

\noindent\textbf{Issue 3: Majority Label.} Finally, we use the TFHE.argmax operation on $\{ \mathcal{n}_j \}_{j=1}^s$ to determine the majority label.
% for the encrypted $\mathbf{x}$. 
% The TFHE.argmax function employs the TFHE.GEQ operation as outlined in Step 4 of Algorithm~\ref{alg:kmeans_argmin}.

\noindent\textbf{(2) Kernel Method.} Kernel evaluation of $k$-NN replaces distance calculations with kernel elements (see Algorithm~\ref{alg:kernel_knn_all}).
%
%% Kernel knn
\begin{algorithm}[!htb]

\DontPrintSemicolon
  % \KwInput{data matrix $\textbf{D} = \{ (\textbf{x}_i, y_i)\}_{i=1}^{n}$, nearest neighbor number $k$, test data $\mathbf{x}$}
  % \KwOutput{predicted class $\hat{y} \in \mathcal{G} =\{\mathcal{g}_1, \dots, \mathcal{g}_s\}$}

  \tcc{parallel compute linear kernel}
  
  % $\textbf{K} = \{ K(\mathbf{x}_i, \mathbf{x}_j) \}_{i, j = 1, \dots, n} $ 

  % $K(\mathbf{x}, \mathbf{x}_i) \leftarrow \mathbf{x}^T \mathbf{x}_i, i=1,\dots,n$

  % \tcc{compute distance $\mathcal{d}_i$}
  $\mathcal{d}_i \leftarrow -2 K(\mathbf{x}_i, \mathbf{x}) + K(\mathbf{x}_i, \mathbf{x}_i), i=1,\dots,n$

  $\bm{\mathcal{d}} \leftarrow (\mathcal{d}_1, \dots, \mathcal{d}_n), \bm{\mathcal{y}} \leftarrow (\mathcal{y}_1, \dots, \mathcal{y}_n)$
  
  % \tcc{find majority class for $\mathbf{x}$}

  $(\mathcal{d}_i^*, l_i^{*})_{i=1,\dots,n} \leftarrow $ TFHE.BubbleSort($\bm{\mathcal{d}}, \bm{y}$)

  $(\mathcal{n}_1, \dots, \mathcal{n}_s)\leftarrow $ TFHE.CountClass($\{l_i^*\}_{i=1,\dots,k}$)

  $\hat{y} \leftarrow$ TFHE.argmax$_{j} \{ \mathcal{n}_j \}_{j=1,\dots,s}$

\caption{TFHE.Kernel$k$-NN($\textbf{D}$, $k$, $\mathbf{x}$)}
\label{alg:kernel_knn_all}
\end{algorithm}

\end{document}